\newcommand*{\PREPRINT}{}

\documentclass{article}

\usepackage{arxiv}

\usepackage[utf8]{inputenc} 
\usepackage{hyperref}       
\usepackage{url}            
\usepackage{booktabs}       
\usepackage{amsfonts}       
\usepackage{amsmath}
\usepackage{graphicx}

\ifdefined\PREPRINT
\else
    \usepackage{lineno}
    \linenumbers
\fi

\title{Latent diffusion models for generative precipitation nowcasting with accurate uncertainty quantification}

\author{
  Jussi Leinonen, Ulrich Hamann, Daniele Nerini, Urs Germann \\
  Federal Office of Meteorology and Climatology MeteoSwiss \\
  Locarno-Monti, Switzerland\\
  \texttt{\{jussi.leinonen,ulrich.hamann,daniele.nerini,urs.germann\}@meteoswiss.ch} \\
  \AND
  Gabriele Franch \\ 
  Fondazione Bruno Kessler \\
  Trento, Italy \\
  \texttt{franch@fbk.eu} \\
}

\begin{document}
\maketitle
\begin{abstract}
Diffusion models have been widely adopted in image generation, producing higher-quality and more diverse samples than generative adversarial networks (GANs). We introduce a latent diffusion model (LDM) for precipitation nowcasting --- short-term forecasting based on the latest observational data. The LDM is more stable and requires less computation to train than GANs, albeit with more computationally expensive generation. We benchmark it against the GAN-based Deep Generative Models of Rainfall (DGMR) and a statistical model, PySTEPS. The LDM produces more accurate precipitation predictions, while the comparisons are more mixed when predicting whether the precipitation exceeds predefined thresholds. The clearest advantage of the LDM is that it generates more diverse predictions than DGMR or PySTEPS. Rank distribution tests indicate that the distribution of samples from the LDM accurately reflects the uncertainty of the predictions. Thus, LDMs are promising for any applications where uncertainty quantification is important, such as weather and climate.
\end{abstract}


\section{Introduction}

Sudden onset of precipitation frequently endangers human lives and causes damage and disruption to infrastructure through flooding and landslides, and is often accompanied by other hazardous weather phenomena such as hail, lightning and windstorms. Precipitation is also a fundamental driver of agriculture and hydroelectric power generation. Consequently, short-term precipitation forecasts are important tools that can benefit infrastructure managers, emergency services and the general public if provided in a timely manner.

Numerical weather prediction (NWP) models can typically forecast the probability and general intensity of precipitation occurring in a wider area, but they struggle at short spatial and temporal scales \cite{Surcel2015ScaleDependence} because of the long running time and the time needed to assimilate data, i.e. to incorporate observational data used as the initial conditions. This problem is particularly severe with convective precipitation, which is associated with the highest rainfall rates, and originates from cells with a spatial scale on the order of a few tens of kilometers, making the exact location of the precipitation difficult to predict with NWP \cite{Sun2014NWPPrecip}. Experience over decades has shown that at lead times of minutes to a few hours, statistical and data-driven models that make optimal use of the latest available observations are useful tools for the short term prediction, or \emph{nowcasting}, of precipitation. Such models have been widely deployed by meteorological agencies.

A common way to implement precipitation nowcasting is \emph{Lagrangian extrapolation}: using motion-detection algorithms to derive motion vectors from consecutive measurements of rainfall by weather radar, then advecting the precipitation field using these vectors to predict its future movement \cite{Bellon1978SHARP,Germann2002ScaleDependence}. The skill of Lagrangian extrapolation decreases rapidly with lead time because of the growth and decay of precipitation, in particular in convective situations. Multiple approaches have been proposed to overcome this limitation, including seamless blending with NWP forecasts (e.g. \cite{Bowler2006STEPS,Sideris2020NowPrecip}) and incorporating information about orographic forcing \cite{Panziera2011NowcastingOrographic,Foresti2018AnalysisOrographic,Sideris2020NowPrecip}. Advanced nowcasting methods also augment the Lagrangian extrapolation framework with features that aim to preserve the structure of precipitation and generate a set of multiple predictions (called an \emph{ensemble} nowcast), where different ensemble members represent possible scenarios of future rainfall and their diversity can be used to quantify the forecast uncertainty. Prominent among such methods is the Short-Term Ensemble Prediction System (STEPS) \cite{Bowler2006STEPS,Seed2013STEPS}, implemented in the PySTEPS open-source library \cite{Pulkkinen2019Pysteps}.

Numerous studies have also used various architectures of deep neural networks (DNNs) for nowcasting (e.g. \cite{Shi2015DeepNowcast,Agrawal2019Nowcast,Ayzel2020RainNet,Franch2020Nowcast}), typically training the network to optimize a metric such as mean squared error (MSE) of the predicted precipitation. DNN-based nowcasting can learn to predict growth and decay, but suffers from blurring of the predictions, where the predicted fields become weaker and more widespread with increasing lead time. This reflects the increasing uncertainty of the prediction resulting from the low predictability of weather. Although such blurred predictions represent the mean expected rainfall, they are not realistic future scenarios. This hinders uncertainty quantification, which is an important aspect of a reliable forecast for downstream applications such as hydrological simulations.

Deep-learning models have also been used to generate more realistic precipitation fields than allowed by simple loss functions, predicting the conditional distribution of the future state of the weather instead of its conditional mean only. This has most often been achieved with Generative Adversarial Networks (GANs; \cite{Goodfellow2020GAN}), which consist of two simultaneously-trained neural networks: a discriminator that is trained to distinguish real samples that belong to the training dataset from generated samples, and a generator that is trained to produce samples that ``fool'' the discriminator, thus learning to produce samples that resemble those in the training set. GANs have been used to create precipitation fields in applications such as postprocessing and downscaling \cite{Leinonen2020Downscaling,Price2022Resolution,Harris2022Downscaling}, precipitation estimation from remote sensing measurements \cite{Hayatbini2019PERSIANNGAN,Wang2021PrecipGAN} and disaggregation \cite{Scher2021Disaggregation}. The state of the art in generative nowcasting is, to our knowledge, presently Deep Generative Models of Rainfall (DGMR) \cite{Ravuri2021GenerativePrecipitation}, which uses a conditional GAN with a regularization term to incentivize the model to produce forecasts close to the true precipitation. DGMR is able to create realistic rainfall predictions that are also numerically accurate, and it can create multiple predictions for each input, enabling ensemble nowcasting.

While GANs are conceptually quite simple, their adversarial training tends to make training them costly and difficult \cite{Mescheder2018GANTraining}. The shifting objectives often cause the convergence to be unstable or slow, and it is necessary to expend training resources to train the discriminator, which is not needed after training in most GAN applications. GANs can also be prone to \emph{mode collapse} \cite{Bau2019ModeCollapse}, where a generator learns to output just one or a few different examples. In conditional GANs this can manifest as the generator ignoring its noise input, always generating identical outputs for a given input.

Denoising diffusion models (DMs), also called score-based generative models, have recently emerged as an alternative to GANs in generative modeling \cite{Song2019GenerativeGradients,Song2020ImprovedScoreBased}. Their mathematical formulation is based on a forward process that gradually degrades an $N$-dimensional sample with increasing amounts of added noise until the sample is indistinguishable from random noise. The neural network is trained to perform one step in an iterative denoising process that reverses the forward process. When the denoising is performed starting from a sample containing only random noise, the reverse process converges to a sample in the training data distribution. DMs have been shown to outperform GANs in terms of sample quality and diversity \cite{Dhariwal2021DiffusionGAN}, and can be conditioned to specific inputs similarly to GANs. In image processing tasks, they have excelled at tasks such as text-to-image generation, inpainting, uncropping and superresolution \cite{Saharia2022Palette,Ramesh2022DALLE2,Saharia2022Imagen,Li2022SRDiff}. DMs are trained to optimize a relatively simple loss function, avoiding the complications of adversarial training and thus making them easier and less computationally expensive to train than GANs. They are also not susceptible to mode collapse. A downside of DMs compared to GANs is the higher cost of generation: since the reverse diffusion process is iterative, the model has to be evaluated several times. Early DMs such as Denoising Diffusion Probabilistic Models (DDPM \cite{Ho2020DDPM}) could require thousands of iterations; this was brought down by alternate process models such as the Denoising Diffusion Implicit Models (DDIM \cite{Song2021DDIM}) to the order of $100$ iterations. Recently, samplers based on pseudo-linear multistep (PLMS \cite{Liu2022PLMS}) differential equation solvers have decreased the number of required iterations further, producing good samples with $30$--$50$ iterations and acceptable ones with as few as $10$.

The ability of DMs to generate diverse samples suggests that they are potentially useful in applications where modeling the uncertainty of predictions is important, such as weather, climate and hydrology. The ability of DMs to generate precipitation fields was recently demonstrated \cite{Addison2022Emulation}. In this work, we introduce the use of DMs for ensemble precipitation nowcasting. To reduce the computational cost, we utilize the latent diffusion model (LDM) concept used by Stable Diffusion \cite{Rombach2022LatentDiffusion}, where the diffusion process is run in a latent variable space mapped to the physical pixel space by an autoencoder. There are three main components of the model, which we call LDCast: 
\begin{enumerate} 
\item \textbf{Forecaster stack}: To condition the model, we introduce a novel spatiotemporal prediction architecture based on Adaptive Fourier Neural Operators (AFNOs) \cite{Guibas2022AFNO,Pathak2022FourCastNet}, with temporal cross attention to map between the input and output time coordinates.
\item \textbf{Denoiser stack}: We adapt the network used by \cite{Rombach2022LatentDiffusion}, using 3D convolutions to model spatiotemporal differences, and an AFNO-based module used in place of cross attention to couple the network to the conditioning.
\item \textbf{Variational autoencoder} (VAE): We use simple 3D convolutional neural networks (CNNs) as the encoder and the decoder in a VAE with a continuous latent space to reduce the number of data points by a factor of $64$. 
\end{enumerate}
To produce samples of forecast future precipitation, the past precipitation field is first encoded with the encoder part of the VAE. Then, the forecaster is used to produce a prediction of the future precipitation; this prediction is used to condition the denoiser, which is run in a loop with the PLMS sampler \cite{Liu2022PLMS} to produce samples in $50$ iterations. Finally, the predicted latent rainfall field is decoded with the VAE decoder. Further details are given in Sect.~\ref{sect:ldm}.

We observe that LDCast creates predictions of the future evolution of precipitation that are visually realistic and highly consistent with the inputs. We compare the outputs to DGMR and PySTEPS benchmarks using two datasets, described in \ref{sect:datasets}: the test set from the Swiss radar-based precipitation dataset on which the model was trained, and a German dataset that was used for evaluation only, providing a test where both LDCast and DGMR are outside the regions of their respective training datasets. With quantitative ensemble forecast accuracy metrics, LDCast outperforms PySTEPS and DGMR, although DGMR sometimes achieves better scores in forecasting whether the precipitation exceeds given thresholds. The clearest advantage of LDCast is in accurate uncertainty quantification. We show that DGMR produces overconfident predictions, i.e. the ensemble members are too close to each other both quantitatively and in terms of the amount of diversity of precipitation patterns produced, while LDCast achieves a realistic assessment of the uncertainty of the forecast. 

\section{Results}

\begin{figure}
    \centering
    \includegraphics[width=\textwidth]{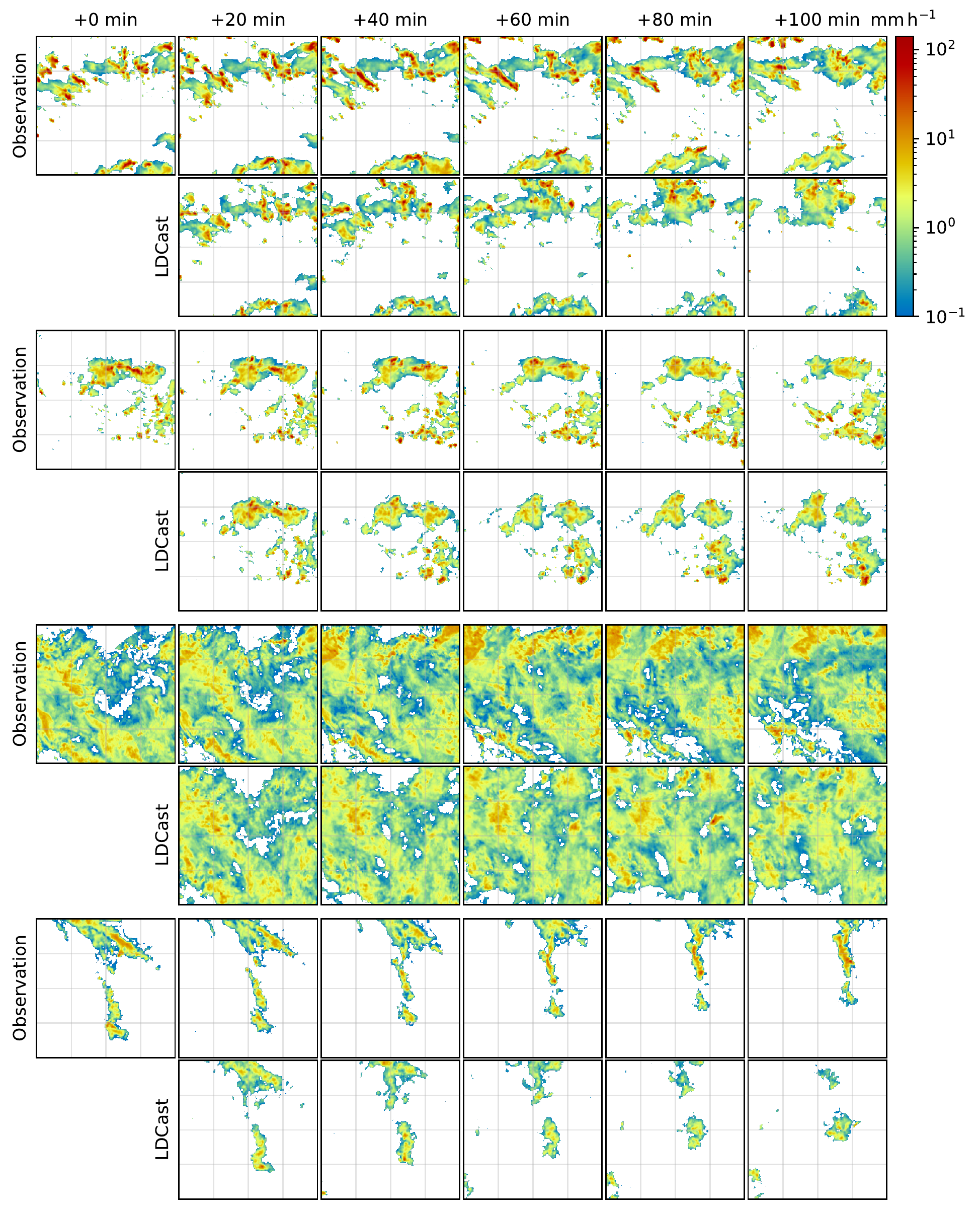}
    \caption{Sample cases of $256\ \mathrm{km} \times 256\ \mathrm{km}$ size comparing the precipitation rate observation and the prediction of the LDCast model. Time steps are produced by the model at $5\ \mathrm{min}$ resolution but they are visualized at $20\ \mathrm{min}$ intervals due to space constraints. The first ensemble member is shown in each case.}
    \label{fig:sample-cases}
\end{figure}

In Fig.~\ref{fig:sample-cases}, we show four examples of precipitation predicted with LDCast. In each case, we show the actual precipitation on the top and one LDCast prediction on the bottom. We display the first ensemble member for each prediction, although any member would be equally valid. The first two cases are from the test set of the Swiss dataset while the last two are from the German dataset.

The first case contains intense convective rainfall. The LDCast prediction contains precipitation cells with a correct intensity and degree of organization, producing line-like structures and clusters similar to those found in the observed precipitation. Not every detail is correctly predicted; however, this cannot be expected at long lead times from a single ensemble member. 

The second case shows an organized convective system at the top and more isolated cells on the bottom left. LDCast again reproduces the correct spatial patterns; the precipitation intensity of the cells on the bottom appears to be roughly correctly predicted while the intensity at the top is somewhat underestimated especially in the $40\ \mathrm{min}$ and $60\ \mathrm{min}$ frames. Interestingly, LDCast correctly predicts the separation of the convective cores on the top, although it forecasts a more complete separation than actually occurs.

The third case shows larger-scale rainfall with embedded convection at more moderate rain rate compared to the first two cases. LDCast again reproduces the degree of spatial organization well and correctly detects the relatively fast motion of the rainfall field from the bottom left towards the top right.

In the fourth case, linear precipitation structures move rapidly towards the top and somewhat towards the right of the images. LDCast correctly predicts the motion and maintains the linear shape until $60\ \mathrm{min}$, after which the predicted rainfall loses cohesion faster than that observed. There is a high variability among the other ensemble members, indicating a low predictability in this case; none of the ensemble members preserve the linear shape quite as strongly as the observation.

In all four cases, it can be seen that the prediction is initially close to the observation and then diverges gradually. This demonstrates that the forecaster stack effectively conditions the prediction to the observed past rainfall.

\subsection{Prediction accuracy}

We used the continuous ranked probability score (CRPS; Sect.~\ref{sect:crps}) as the quantitative metric for assessing the accuracy of the precipitation rate predictions. CRPS takes into account the distribution of the $32$ ensemble members, making it suitable for ensemble forecast verification. CRPS is evaluated pixelwise and thus does not reflect the accuracy of the spatial patterns in the prediction. To assess whether precipitation is correctly predicted over different spatial scales, we also calculated the CRPS for precipitation averaged over $8\ \mathrm{km} \times 8\ \mathrm{km}$ and $64\ \mathrm{km} \times 64\ \mathrm{km}$ windows. Furthermore, to give a metric of the relative error in addition to the absolute error, we computed the CRPS for the logarithm of the rainfall (LogCRPS) using a fill value of $0.02\ \mathrm{mm\,h^{-1}}$ for regions of zero rainfall (as used when training LDCast).

\begin{figure}
    \centering
    \includegraphics[width=\textwidth]{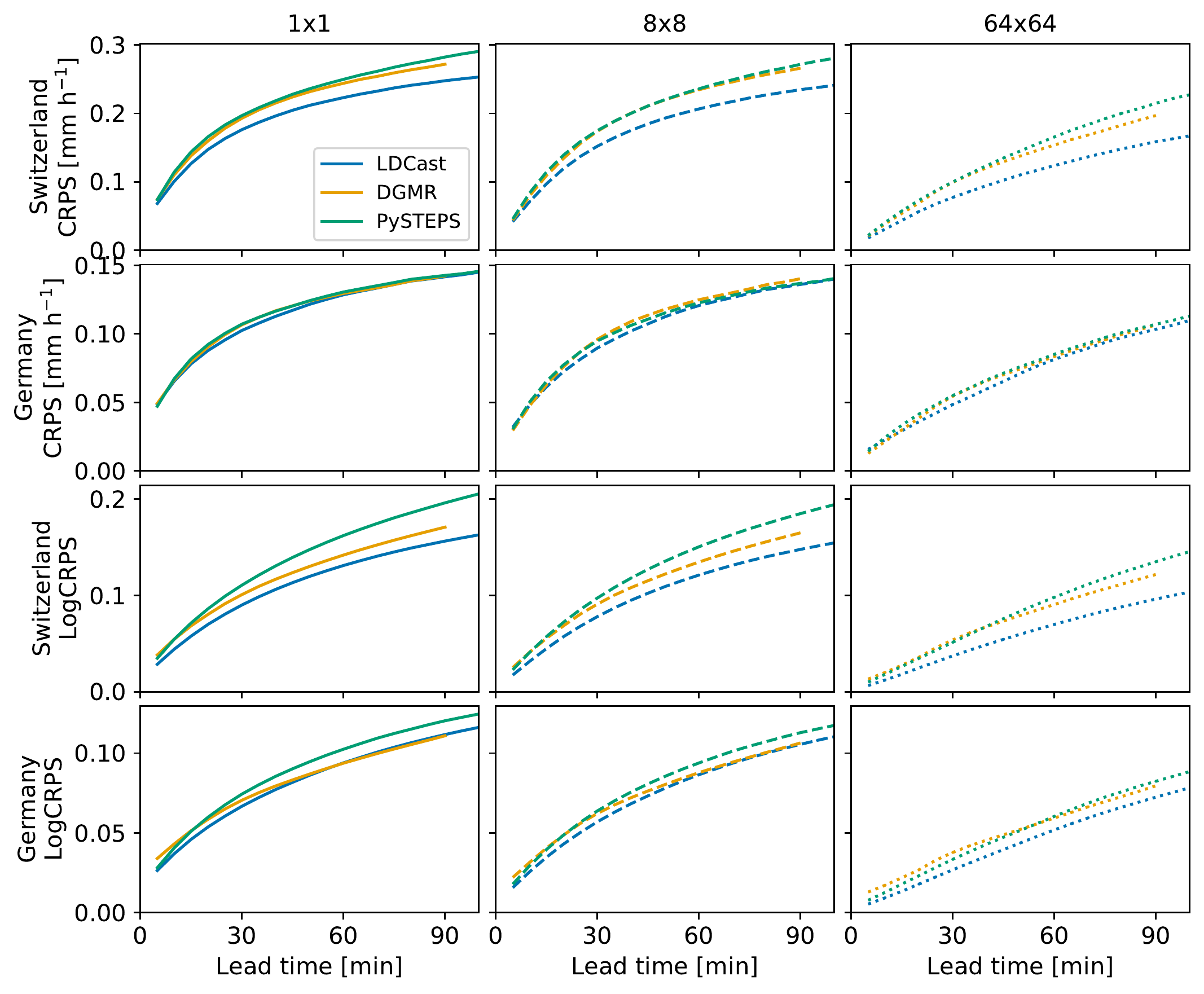}
    \caption{CRPS (lower is better) for the LDCast model as a function of the forecast lead time, compared to the DGMR and PySTEPS benchmarks. The two top rows show CRPS for the absolute precipitation $R$ while the bottom rows show LogCRPS, i.e. CRPS for $\log_{10}(R)$. The three columns correspond to different amounts of averaging: no averaging ($1\ \mathrm{km}$ scale) for the first column, $8\ \mathrm{km} \times 8\ \mathrm{km}$ averaging for the second and $64\ \mathrm{km} \times 64\ \mathrm{km}$ for the third.}
    \label{fig:crps-leadtime}
\end{figure}

The results of the CRPS calculation as a function of lead time are shown in Fig.~\ref{fig:crps-leadtime}. The CRPS from LDCast is compared to DGMR and PySTEPS ensemble predictions. With the Swiss dataset, LDCast clearly outperforms DGMR and PySTEPS at all scales in both CRPS and LogCRPS. The advantage of LDCast over the other models increases with longer lead times. With the German dataset, all three models are quite close to each other in CRPS, with LDCast achieving slightly better overall scores. There are somewhat larger differences between the models in LogCRPS of the German dataset, with LDCast the best model in most situations. 

\subsection{Representation of uncertainty}

\begin{figure}
    \centering
    \includegraphics[width=\textwidth]{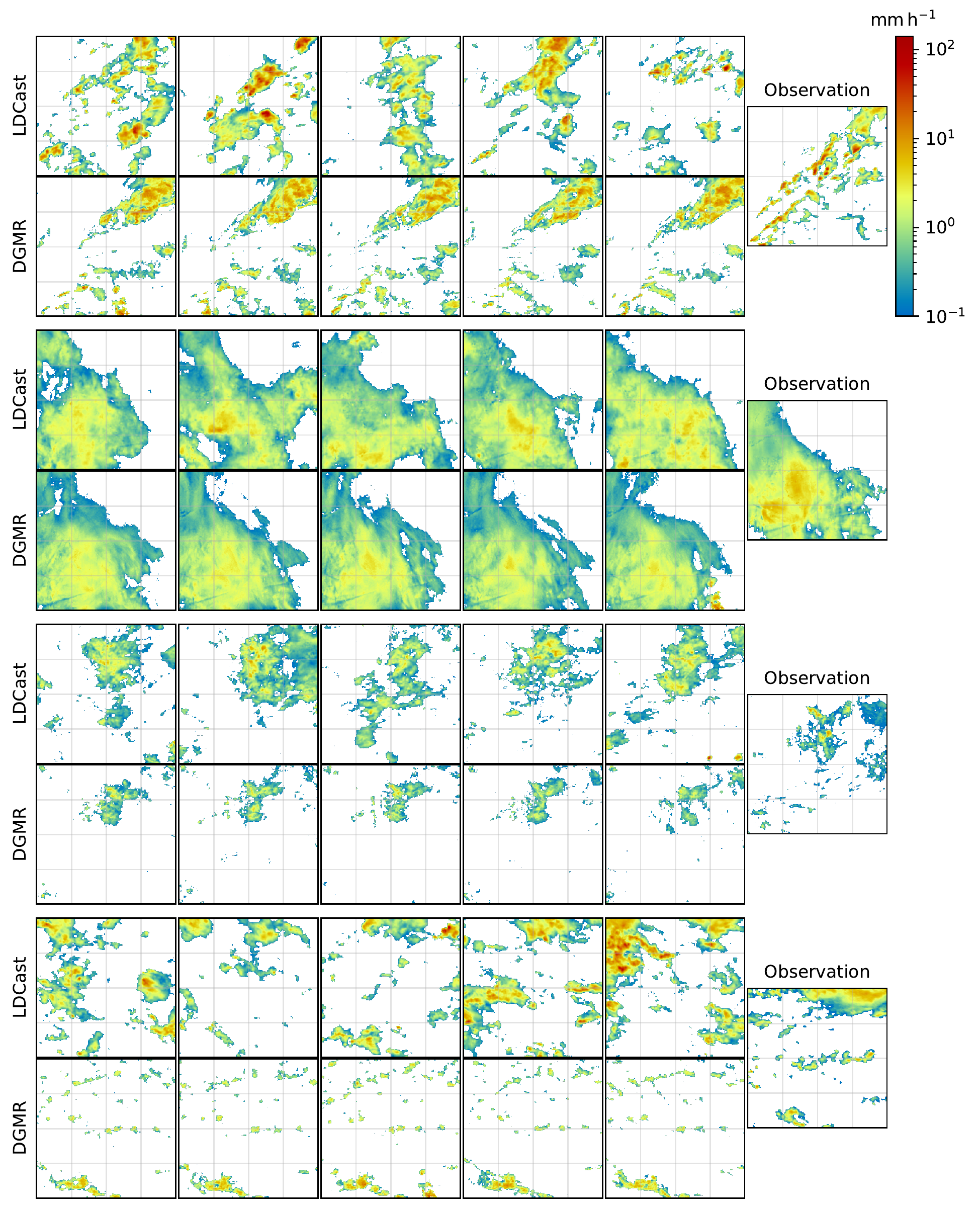}
    \caption{Ensemble members of predicted precipitation at $90\ \mathrm{min}$ lead time. In each of four cases, the results from LDCast are shown on the first row on the left and the results from DGMR on the second row. The actual observed precipitation is shown for comparison on the right.}
    \label{fig:ensemble-diversity}
\end{figure}

In Fig.~\ref{fig:ensemble-diversity}, we show examples of the first five ensemble members of LDCast and DGMR at $90\ \mathrm{min}$ lead time. This is the maximum lead time of DGMR, and thus the prediction where the largest variability between ensemble members is expected. As with Fig.~\ref{fig:sample-cases}, the first two examples are from the Swiss test dataset while the last two are from the German dataset. Visual comparison of the LDCast and DGMR outputs shows that the DGMR ensemble members are rather similar to each other, while the variability of the LDCast outputs is much greater. Notably, the mutual similarity of the DGMR outputs appears greater than their similarity to the observation, suggesting that DGMR produces overconfident predictions.

\begin{figure}
    \centering
    \includegraphics[width=\textwidth]{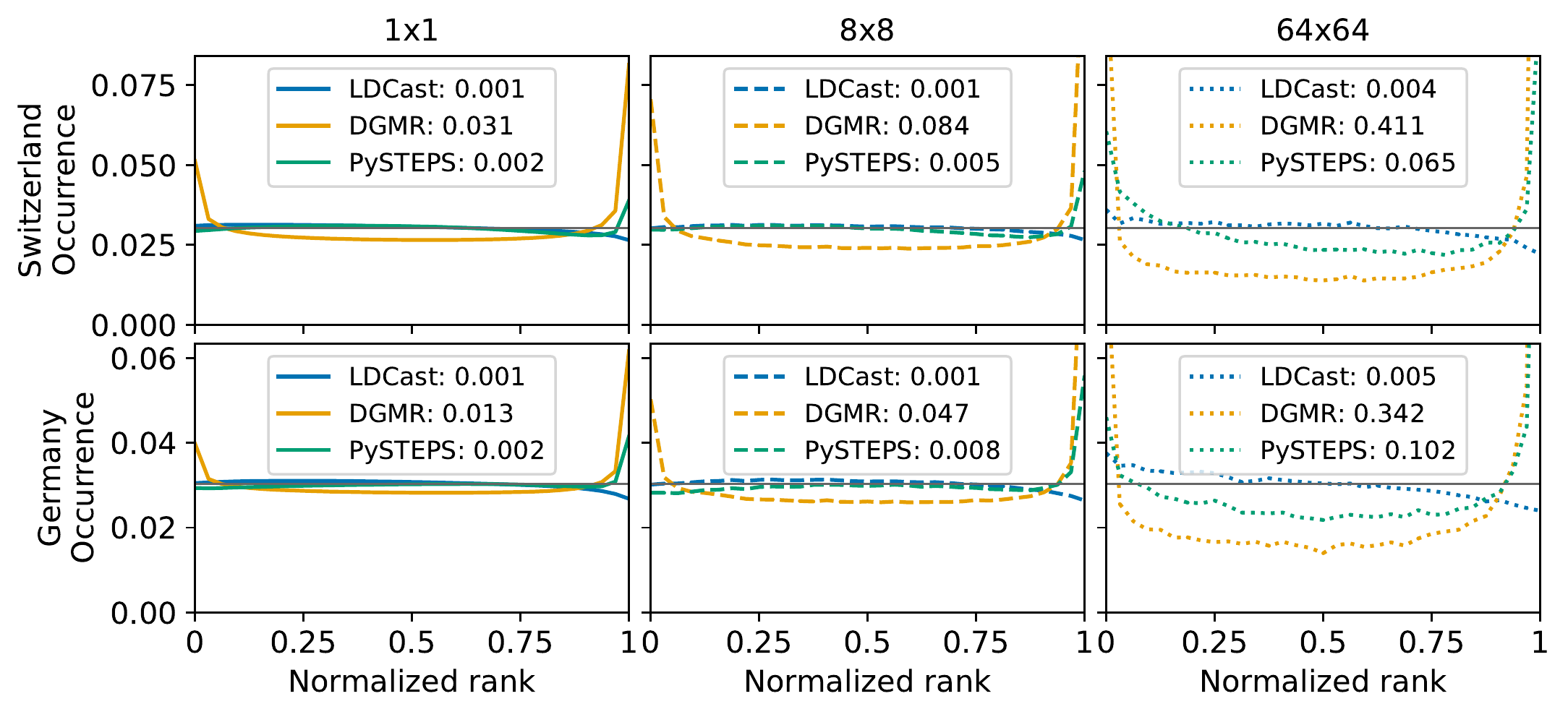}
    \caption{Rank distributions for the LDCast, DGMR and PySTEPS models. The columns correspond to different averaging scales as with Fig.~\ref{fig:crps-leadtime}. The numbers in the legend indicate the Kullback--Leibler divergence from the uniform distribution. The gray line in each plot indicates the ideal uniform distribution.}
    \label{fig:rank-distribution}
\end{figure}

We can quantitatively examine the correctness of the uncertainty estimates using rank distributions (Sect.~\ref{sect:rank}). These are shown in Fig.~\ref{fig:rank-distribution} for multiple scales and compared to DGMR and PySTEPS. Similar to Fig.~\ref{fig:crps-leadtime}, we also show the results for rainfall averaged over $8\ \mathrm{km} \times 8\ \mathrm{km}$ and $64\ \mathrm{km} \times 64\ \mathrm{km}$ windows. The LDCast results are closest to the ideal flat distributions. DGMR rank distributions are ``U-shaped'', that is, they contain many high and low ranks, corresponding to overconfident predictions in agreement with the qualitative comparison above. PySTEPS rank distributions at the $1\ \mathrm{km} \times 1\ \mathrm{km}$ and $8\ \mathrm{km} \times 8\ \mathrm{km}$ scales contain too many high ranks (but not too many low ones), indicating that PySTEPS produces many cases where all ensemble members underestimate the precipitation. At the $64\ \mathrm{km} \times 64\ \mathrm{km}$ scale PySTEPS also produces a U-shaped distribution, while that of LDCast is still relatively flat. The Kullback--Leibler divergence (KL) from the uniform distribution shows that LDCast achieves scores clearly closest to the optimum. The rank distribution results are very similar between the Swiss and German datasets.

\subsection{Forecasting event occurrence}

\begin{figure}
    \centering
    \includegraphics[width=\textwidth]{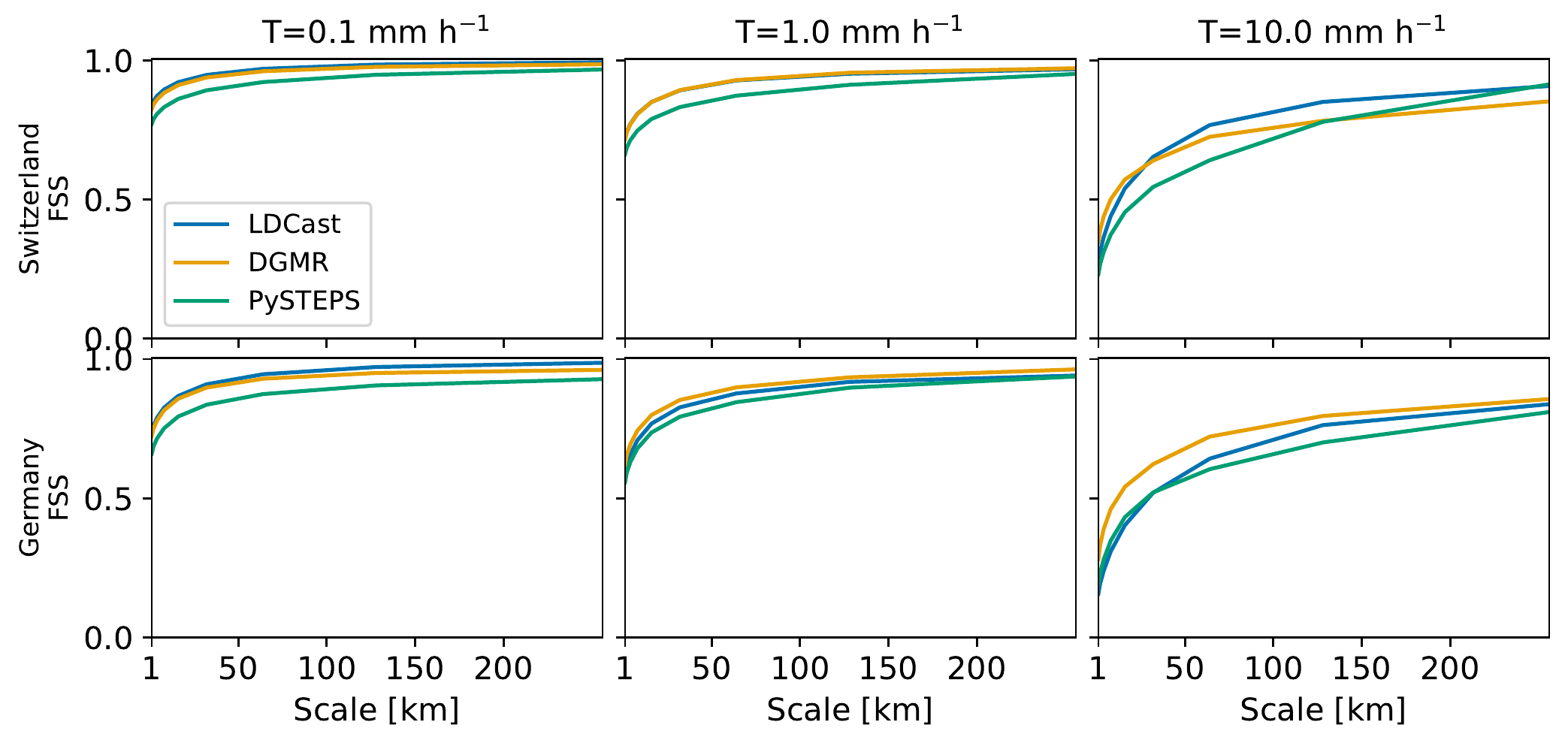}
    \caption{FSS as a function of scale for LDCast, DGMR and PySTEPS. The three columns show the FSS for thresholds of $0.1\ \mathrm{mm\,h^{-1}}$, $1.0\ \mathrm{mm\,h^{-1}}$ and $10.0\ \mathrm{mm\,h^{-1}}$, respectively.}
    \label{fig:fss-threshold}
\end{figure}

We used the fractions skill score (FSS; Sect.~\ref{sect:fss}) to measure the skill of the models at predicting whether the precipitation exceeds certain threshold values. We computed the FSS at scales of $2^N\ \mathrm{km}$ (with $N$ an integer) up to $256\ \mathrm{km}$. The results are shown in Fig.~\ref{fig:fss-threshold} for thresholds of $0.1\ \mathrm{mm\,h^{-1}}$, $1\ \mathrm{mm\,h^{-1}}$ and $10\ \mathrm{mm\,h^{-1}}$, averaged over all lead times. With the Swiss test dataset, LDCast performs approximately equally to DGMR and better than PySTEPS at all scales for the $R \geq 0.1\ \mathrm{mm\,h^{-1}}$ and $R \geq 1\ \mathrm{mm\,h^{-1}}$ thresholds; for $R \geq 10\ \mathrm{mm\,h^{-1}}$, the results are similar except LDCast achieves better scores at the $32$--$128\ \mathrm{km}$ scales. With the German dataset, LDCast is slightly better than DGMR at the $0.1\ \mathrm{mm\,h^{-1}}$ threshold, while being slightly behind at $1\ \mathrm{mm\,h^{-1}}$ and considerably behind at $10\ \mathrm{mm\,h^{-1}}$. The generative models based on deep learning perform better than PySTEPS in all cases except $R \geq 10\ \mathrm{mm\,h^{-1}}$ at long scales for the Swiss dataset and $R \geq 10\ \mathrm{mm\,h^{-1}}$ at short scales for the German dataset.

\section{Discussion}

In this article, we have introduced the use of latent diffusion models for generative nowcasting of precipitation measured by weather radars. Our model, LDCast, generates ensembles of realistic precipitation fields, using $4$ time steps ($20\ \mathrm{min}$) of precipitation as its input, and predicting precipitation up to $20$ time steps ($100\ \mathrm{min}$) to the future. Quantitative comparisons to DGMR, a GAN-based precipitation nowcasting model, and to PySTEPS, a commonly used statistical nowcasting algorithm, reveal that LDCast outperforms them in accuracy (measured by CRPS). LDCast has a particularly distinct advantage over the benchmark models in characterizing the uncertainty of its predictions, generating diverse forecasts that result in rank distributions that are much closer to uniform. This diversity makes it easier for the model to reveal the possibility of less likely but higher impact events, such as extreme weather. The advantage of LDCast over the benchmark models increases when precipitation averaged over a larger scale is considered. Meanwhile, the results are more mixed with regard to the ability of the models to predict whether the precipitation exceeds predetermined thresholds, as measured by the FSS. A possible factor in this is that LDCast was trained on a logarithmic transformation of the precipitation rate, thus emphasizing the relative error, while DGMR was trained directly with the precipitation rate, which can be expected to emphasize the absolute error.

We evaluated the models using two different test datasets. One was from Switzerland and its surroundings, the same region where the model was trained. To assess how well the model generalizes to outside its training domain, we also performed the evaluation with rain rate data from northern Germany. The comparisons to the benchmark models indicate that LDCast loses some of its advantage over DGMR in CRPS and FSS when evaluated in the out-of-domain dataset. One reason for this may be that northern Germany and United Kingdom, from where the DGMR training data were obtained, are at similar latitudes and in proximity of the North Sea, and therefore have climates that resemble each other more closely than that of Switzerland, which experiences more convective precipitation, and where the evolution of precipitation patterns is expected to be different due to the orographic influence of the Alps. In contrast to FSS, LDCast retains its superiority in the rank histograms also with the German dataset. Thus, its ability to quantify its own uncertainty appears to be quite robust.

Another advantage of LDMs is the relative ease of training compared to GANs. On our system of eight Nvidia V100 GPUs, we initially trained our model for approximately $53\ \mathrm{h}$ with $128 \times 128$ pixel samples, then fine tuned it for approximately $5\ \mathrm{h}$ with $256 \times 256$ pixel samples. These computational costs, while significant, are considerably lower compared to GAN-based models. For comparison, we briefly experimented with implementing DGMR from the pseudocode available in \cite{Ravuri2021GenerativePrecipitation} (the full source code for DGMR is not available; for the DGMR benchmark, we used the saved generator released by the developers). The training speed that we achieved indicated that training for the full $5 \times 10^5$ generator steps described in \cite{Ravuri2021GenerativePrecipitation} would have required approximately $1100\ \mathrm{h}$, i.e. $46$ days, on the abovementioned hardware. As this was not critical to our investigation, we decided to forgo training the model to completion. Optimized implementations might improve the training times for both models; nevertheless, it seems clear that LDMs make generative modeling in weather and climate sciences more approachable to researchers with limited computational resources. Beyond training speed, our experience with training the model was also that the stability of training diffusion models makes the development process easier compared to GANs. Furthermore, compared to our initial attempts to generate samples in the pixel space, we found that the latent-space encoding in LDMs not only reduces computational costs but also improves training stability by regularizing the input and output variable space.

A downside of DMs is that the network needs to be evaluated several times during sample generation. Our sampling process required approximately $19\ \mathrm{s}$ to generate one $20 \times 256 \times 256$ ensemble member on one of the GPUs used for training. This can potentially be reduced in operational use by using fewer sampler iterations (possibly at the cost of lower sample quality and diversity), with lower precision floating point arithmetic, and with architectural and implementation optimizations to minimize redundant calculations between iterations. Ensemble members can also be generated in parallel with multiple GPUs. Nevertheless, the computational requirements make DMs less likely to be adopted in performance-critical applications, such as using neural networks to emulate computationally expensive components of weather and climate models. For such applications, latent-space autoencoders can also be used in combination with a GAN \cite{Esser2021Taming}, which may provide performance benefits. Another limitation of the iterative nature of DMs is that as implicit models, it is not straightforward to include physics-based or statistical constraints in them. Further research is needed to determine how such constraints could be implemented in DMs. Nevertheless, LDCast performs well compared to DGMR, which does include statistical constraints on the generated precipitation, implying that such constraints are not necessarily needed in practice.

Precipitation nowcasting has for several years drawn considerable attention as an application of deep learning. However, nowcasting turned out to be a challenging application for deep generative models, and appeared relatively late with the recent introduction of models like DGMR. The success of LDMs at this task, combined with the computational advantages, suggests that they will find applications in nowcasting different atmospheric variables, as well as in other weather and climate applications in which accurate uncertainty quantification is important. We also expect that the LDCast methodology can be extended to exploit multiple predictor variables, potentially including satellite observations and forecasts from numerical weather prediction models similar to \cite{Leinonen2022Lightning}. Our forecaster stack based on AFNO and temporal attention with positional encoding is naturally suited for this as it can flexibly handle inputs at different time coordinates.

\section{Methods} \label{sect:methods}

\subsection{Datasets} \label{sect:datasets}

We trained the model on a dataset of precipitation rate estimates from the MeteoSwiss operational radar network \cite{Germann2006RadarMountainous,Germann2016SwissRadar}. The network consists of five scanning C-band Doppler radars, whose overlapping ranges, optimized scanning strategy and processing algorithms (vertical profile, visibility and clutter correction) mitigate the issue of topographic blocking in the complex Swiss terrain. The radar composite is produced every $5\ \mathrm{min}$ at $1\ \mathrm{km}$ resolution in a rectangular area $710\ \mathrm{km}$ in the east--west direction and $640\ \mathrm{km}$ north--south, covering all of Switzerland and some surrounding regions. The data were gathered from the years 2018--2021, using the period from April to September for each year to focus the training more on the convective season, when the variability of rain rates is largest.

In order to test models outside the region in which they were trained, we also obtained precipitation rate data from the radar composite of the German Weather Service (DWD) \cite{Stephan2008DWDRadar} from April--September 2022. This network covers all of Germany, but the southern part partially overlaps with the Swiss radar network, so we only use the northern half for testing. The $5\ \mathrm{min}$ / $1\ \mathrm{km}$ temporal and spatial resolutions of the DWD and MeteoSwiss composites are identical to each other, and also to those of the UK MetOffice radar network, which was used to train DGMR. Thus, both LDCast and DGMR can be evaluated without retraining in both the Swiss and German domains.

We split the Swiss dataset to training, validation and testing sets such that each UTC day is assigned entirely to only one of the splits; this is done to reduce the temporal proximity, and hence correlation, of the training and validation/testing data. Approximately $10\%$ of the data is assigned to the validation set and another approximately $10\%$ to the testing set. The German dataset is used only for testing. The final evaluation is performed with $1024$ samples from each testing dataset, with $32$ ensemble members generated for each sample with each model.

When generating training, validation and testing samples, rather than sampling the datasets uniformly we sample them such that the model sees similar numbers of cases from different precipitation intensities $R$. This is achieved by oversampling cases containing higher $R$. We divide the dataset into $32 \times 32$ pixel tiles, and compute $R_\mathrm{m}$, the 99th percentile of precipitation rate in each tile (representing a soft maximum less sensitive to outliers). Each tile is then assigned to one of $11$ bins, where the first bin is for $R_\mathrm{m} < 0.2\ \mathrm{mm\,h^{-1}}$, the last bin for $R_\mathrm{m} \geq 50\ \mathrm{mm\,h^{-1}}$, and the rest are logarithmically spaced between $0.2$--$50\ \mathrm{mm\,h^{-1}}$. Training samples are then generated such that each bin is sampled with equal probability.

For preprocessing we follow the strategy of \cite{Leinonen2020Downscaling}. Before feeding samples to the model, they are preprocessed with a logarithmic transformation
\begin{equation}
    f(R) = \begin{cases}
        \log_{10} R & R \geq 0.1\ \mathrm{mm\,h^{-1}} \\
        \log_{10} 0.02 & R < 0.1\ \mathrm{mm\,h^{-1}}
    \end{cases}
\end{equation}
The discontinuity at $0.1\ \mathrm{mm\,h^{-1}}$ is useful for giving the model a clearer distinction between the raining and non-raining points, but we found it could create artifacts in generative models. To mitigate this and other artifacts in the input data, we further apply antialiasing to the samples with a Gaussian filter of $0.5$ pixel standard deviation.

\subsection{Latent diffusion model} \label{sect:ldm}

\begin{figure}
    \centering
    \includegraphics[width=\textwidth]{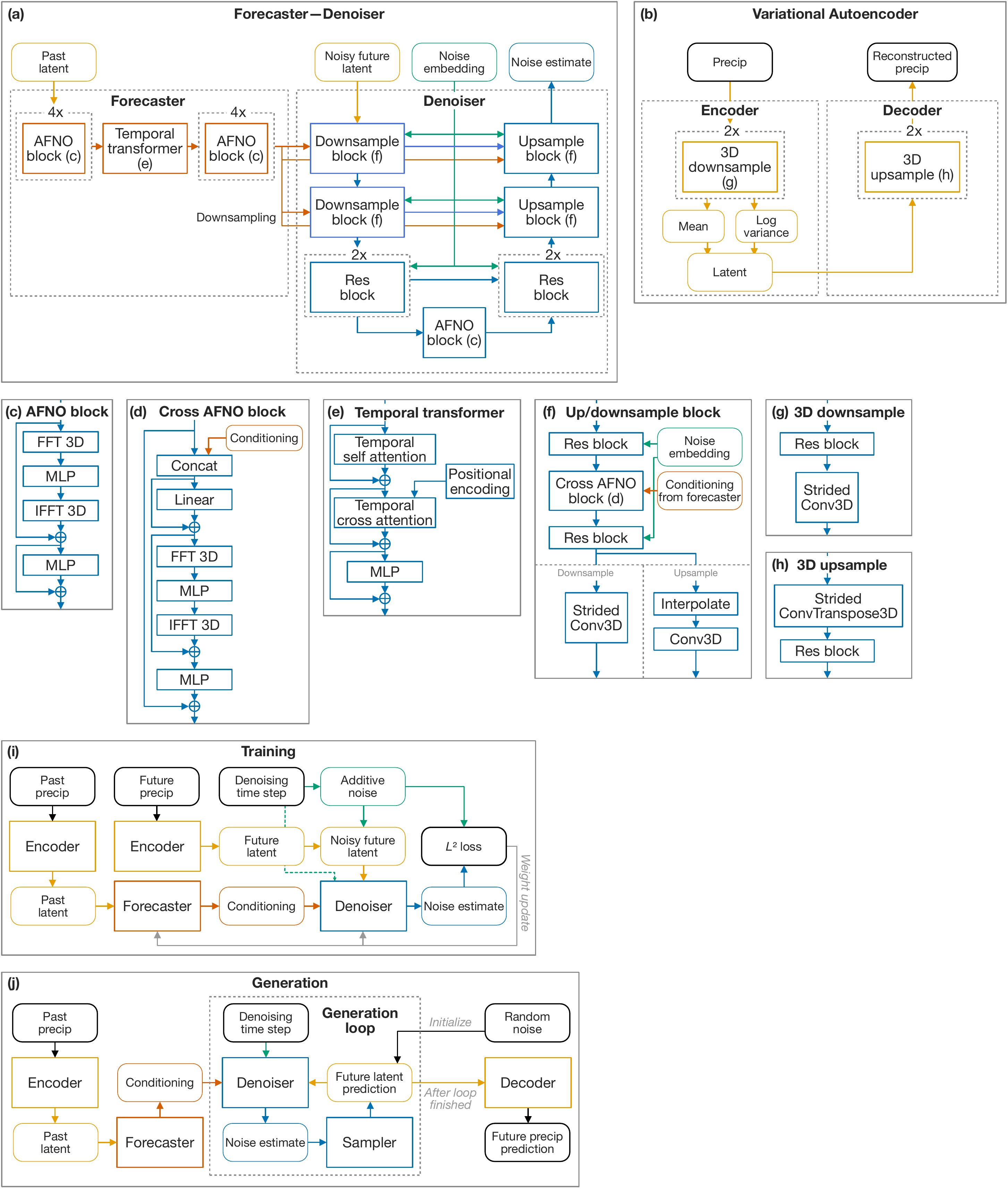}
    \caption{An overview of the LDCast neural networks. (a) The forecaster and denoiser stacks. (b) The VAE used to transform precipitation sequences to the latent space. (c)--(h) The layer blocks used in the network diagrams. (i) The training procedure. (j) The forecast generation procedure.  ``Conv'' denotes convolution. ``MLP'' (multilayer perceptron) is a block consisting of a linear layer, activation function and another linear layer. ``Res block`` denotes a ResNet-type residual block \cite{He2016ResNet}; the noise embedding is added to the input of the block.}
    \label{fig:ldcast-networks}
\end{figure}

LDCast is a conditional LDM that consists of three main network components: a forecaster stack, a denoiser stack and a variational autoencoder. An overview of the network structure is shown in Fig.~\ref{fig:ldcast-networks}. Below, we describe the components of the network and the training process. Implementation details such as hyperparameters can be found in Supplementary Information Table S1. The exact information can be found in the published code as indicated under \hyperref[sect:codeavailability]{Code Availability}.

\subsubsection{Forecaster}

The forecaster stack is based on the AFNO. In the FourCastNet architecture \cite{Pathak2022FourCastNet}, a series of 2D AFNO blocks is used to process the atmospheric state at time step $t$ to predict the state at $t+1$. Each block consists of an AFNO and a pixelwise multilayer perceptron (MLP) network. The model is initially trained to predict one time step, then fine tuned to predict two time steps, and can then be evaluated iteratively to predict further time steps. We modify this procedure for the nowcasting application, where we want to train the model to predict $D_\mathrm{out}=5$ encoded output time steps simultaneously from $D_\mathrm{in}=1$ encoded input time steps. The modified architecture consists of three stages:
\begin{enumerate}
    \item \textbf{Analysis}: The input of dimension $C \times D_\mathrm{in} \times W \times H$, where $C$, $W$ and $H$ are the number of latent-space channels, width and height of the encoded input respectively, is processed with a series of AFNO+MLP blocks.
    \item \textbf{Temporal transformer}: The input is projected to the $C \times D_\mathrm{out} \times W \times H$ output space using a cross-attention transformer \cite{Vaswani2017Attention} block that is only evaluated along the temporal dimension. The query of the cross attention is computed from sinusoidal positional encoding (as in \cite{Vaswani2017Attention}) of the time coordinates of the outputs.
    \item \textbf{Forecast}: The forecast stage is identical in architecture to the analysis stage, but operates in the output space.
\end{enumerate}

We note that this architecture can be used on its own for non-generative prediction. We also expect (although we do not utilize this capability in the current study) that the cross-attention mapping can be used naturally with inputs that have a variable time difference to the outputs and/or a different time resolution compared to the outputs. This adds to the flexibility of the architecture compared to the convolutional and recurrent-convolutional networks that have been frequently used for precipitation nowcasting (e.g. \cite{Franch2020Nowcast,Leinonen2023ThunderstormHazard}).

\subsubsection{Denoiser}

Our denoising stack is a modification of the U-Net-type network used by the original LDM implementation \cite{Rombach2022LatentDiffusion}. To model spatiotemporal relationships, we replaced the 2D convolutions with 3D convolutions. We removed the spatial attention layers of the original network since they add considerable computational cost and removing them did not seem to degrade performance; this is likely due to the spatiotemporally equivariant nature of our data. Furthermore, we noticed that when using the layer normalization employed in the original LDM network, LDCast often produced outputs with realistic spatial patterns but a shifted magnitude of the precipitation intensity. A simple solution was to remove the normalization layers; this allowed the model to reproduce the intensity of the rainfall better, and did not seem to impede convergence significantly.

To condition the denoising network with the forecasting network, we use blocks that concatenate the U-Net state to the conditioning variable, then apply an AFNO operation similar to that used in the forecaster to the concatenated input (Fig.~\ref{fig:ldcast-networks}d). This is based on the reasoning of \cite{Guibas2022AFNO} that AFNO is used in a manner analogous to self-attention; we thus aim at a cross attention-like operation with this block.

\subsubsection{Variational autoencoder}

The VAE is used to encode samples from the pixel space to a continuous latent space and then decode them back to the pixel space. We construct the encoder and decoder parts of the VAE as simple 3D convolutional networks, where each level consists of a ResNet-type residual block and a downsampling (encoder) or upsampling (decoder) convolutional layer. Each level reduces each spatial and temporal dimension by a factor of $2$; we use two levels to reduce the number of points by a factor of $4 \times 4 \times 4 = 64$. The encoder output is bottlenecked to $32$ channels. Between the encoder and decoder stages, the VAE latent space is regularized with a loss based on Kullback--Leibler divergence (KL) between the latent variable and a multivariate standard normal variable.

While the number of spatiotemporal grid points is reduced by a number of $64$ in the encoding process, the number of channels is also increased from $1$ to $32$. Thus, the total amount of data is decreased only by a factor of $2$. However, we found that the reduction in spatial resolution is more important for reducing the computational cost of the forecaster and denoiser stacks. Therefore, the performance gain obtained by operating in the latent space is considerably larger than the data reduction factor.

\subsubsection{Training}

The VAE was trained before the rest of the network, using $L^1$ loss and the KL regularization term. Once trained, the VAE weights were held fixed while the forecaster and denoiser stacks were trained simultaneously. The model was trained to predict $5$ time steps from $1$ time step in the latent space, corresponding to predicting $20$ time steps ($100\ \mathrm{min}$) from $4$ time steps ($20\ \mathrm{min}$) in the pixel space. 

The conditional LDM training loss can be parameterized as an $L^2$ loss \cite{Rombach2022LatentDiffusion}
\begin{equation}
L_\mathrm{LDM} = \mathbb{E}_{\mathcal{E}(x),y,\epsilon \sim \mathcal{N}(0,\mathbf{I}),t} \left [ \Vert \epsilon - \epsilon_\theta(z_t,t,\tau_\theta(y)) \Vert^2_2 \right ]
\end{equation}
where $x$ is the condition (past precipitation), $\mathcal{E}$ is the encoder, $y$ is the real sample (observed future precipitation), $\epsilon$ is random noise, $t$ is the step of the denoising process, $z_t$ is the noisy latent-space sample at step $t$, $\tau_\theta$ is the conditioning (forecaster) stack, $\epsilon_\theta$ is the denoiser and $\theta$ represents the trainable parameters of the networks.

We used the AdamW optimizer \cite{Loshchilov2018AdamW} to train both networks. The hyperparameters are found in Supplementary Information Table S1. The learning rate schedule was based on monitoring the loss in the validation set after every checkpoint, which were performed every $1000$ training batches; if the validation loss did not decrease for a $3$ consecutive checkpoints, the learning rate was reduced. Early stopping was also used, terminating the training after $6$ checkpoints had passed without improvement in the validation loss. Exponential moving averaging (EMA) was applied to the network weights, following \cite{Rombach2022LatentDiffusion}.

The model was initially trained to convergence with $128 \times 128$ pixel samples to reduce training time. It was then fine-tuned with another training run, in which the model was initialized with the weights obtained in the pre-training, using $256 \times 256$ pixel samples. This saves considerable training time compared to training the model from random initialization with $256 \times 256$ pixel samples.

\subsubsection{Evaluation}

We produced samples using the standard LDM approach (Fig.~\ref{fig:ldcast-networks}j):
\begin{enumerate}
    \item The input precipitation is encoded to the latent space using the VAE encoder.
    \item A prediction is computed from the latent-space inputs using the forecaster stack.
    \item Starting from $\mathcal{N}(0,\mathbf{I})$ distributed random noise, we perform $50$ iterations of the denoiser with the PLMS sampler, using the prediction obtained from the previous step for conditioning.
    \item The denoised latent variables thus obtained are decoded to precipitation using the VAE decoder.
\end{enumerate}
The AFNO layers operate similarly to fully convolutional layers, so the model can be trained with samples of one size and then applied to another. We performed the evaluation with $256 \times 256$ pixel samples that were also used in the fine-tuning phase of the training. We also experimented with evaluating the model trained only with $128 \times 128$ pixel samples; the results were quite similar to those for the final model, suggesting that the fine tuning may be omitted if desirable from a computational perspective.

\subsubsection{Postprocessing}

When using the LDCast output, we set all precipitation rate predictions below $0.1\ \mathrm{mm\,h^{-1}}$ to zero and cap the precipitation rate to the maximum in the Swiss dataset, approximately $118\ \mathrm{mm\,h^{-1}}$. When computing quantitative scores (CRPS, FSS and the rank histograms) for LDCast, we reduce bias by using probability matching (PM) based on results on the validation set of the Swiss dataset. That is, we compute the cumulative distribution functions (CDFs) of the predicted values and observed values on the Swiss validation set, and then apply adjustments to the predictions such that the CDFs match. The PM based on the Swiss validation set is used to adjust both the results for the Swiss test set and those for the German dataset. In order to compare the models fairly, we use the same postprocessing procedure also for the benchmarks.

\subsection{Verification scores} \label{sect:metrics}

\subsubsection{Continuous ranked probability score} \label{sect:crps}

The CRPS \cite{Gneiting2007ProperScoring} measures the accuracy of a probabilistic forecast, taking into account both the bias and the spread. Using $i$ to denote a single point in a multidimensional dataset, let $y_i$ be the observation at that point and $\hat F_i$ be the CDF of corresponding probabilistic forecast $\hat y_i$. The CRPS at $i$ is defined as the integral of the squared difference of $\hat F_i$ and the CDF of $y_i$, a unit step function $H$:
\begin{equation}
    \mathrm{CRPS}(\hat F_i,y_i) = \int_{-\infty}^\infty \left ( \hat F_i(x) - H(x-y_i) \right )^2 \mathrm{d}x
\end{equation}
where
\begin{equation}
    H(x) = \begin{cases}
        0 & x \leq 0 \\
        1 & x > 0 
    \end{cases} 
\end{equation}

When an ensemble is used to represent the probability of the forecast, there are $N_\mathrm{e}$ discrete forecasts at $i$: $\hat y_{i,1},\ldots,\hat y_{i,{N_\mathrm{e}}}$. The forecast CDF is then a function consisting of multiple steps:
\begin{equation}
\hat F_i(x) = \frac{1}{N_\mathrm{e}} \sum_{k=1}^{N_\mathrm{e}} H(x-\hat y_{i,k}).
\end{equation}

The CRPS for an entire dataset (or a subset of it) of $N_\mathrm{s}$ samples is computed as the average $N_\mathrm{s}^{-1} \sum_{i=1}^{N_\mathrm{s}} \mathrm{CRPS}(\hat F_i,y_i)$. In the special case of $N_\mathrm{e}=1$, the CRPS over a dataset is simply the mean absolute error (MAE) between the forecast and the observation. Thus, CRPS can be viewed as a generalization of the MAE for probabilistic forecasts.

\subsubsection{Probability integral transform / rank distribution} \label{sect:rank}

The probability integral transform (PIT) tests whether a probabilistic prediction has the same probability distribution as the observations, that is, whether the uncertainty of the predictions is modeled correctly. 

Using the notation adopted in Sect. \ref{sect:crps}, we first define at each point $i$
\begin{equation}
r_i = \hat F_i (y_i),
\end{equation}
that is, $r_i \in [0,1]$ is the value of the forecast CDF at the observation. PIT is based on the fact that if $y$ and $\hat y$ come from the same distribution, the distribution of $r_i$ over the dataset approaches the standard uniform distribution $U_{[0,1]}$ as $N_\mathrm{s} \rightarrow \infty$.

By computing $r$ over a dataset, one can examine the uniformity of the resulting distribution $p_r$. This can be done either visually by plotting the distribution, or quantitatively by computing a distribution distance metric between $p_r$ and the standard uniform distribution $U_{[0,1]}$. One possible metric is the Kullback--Leibler divergence (KL) frequently used in machine learning.

In the case of ensemble forecasts, the PIT is equivalent to the \emph{rank distribution} (or rank histogram) \cite{Candille2006PredictionEvaluation} frequently used in ensemble forecast verification. In this case, $r_i$ is equivalent to the rank of the observation among the forecasts (i.e. the number of forecasts that are smaller than the observation; ties are randomized) divided by the number of ensemble members $N_\mathrm{e}$:
\begin{equation}
r_i = \frac{1}{N_\mathrm{e}} \sum_{k=1}^{N_\mathrm{e}} H(y_i-\hat y_{i,k})
\end{equation}
Consequently, the distribution $p_r$ is discrete, with possible values $r_j = j/N_\mathrm{e}$, $j \in {0 \ldots N_\mathrm{e}}$. One should thus use the discrete version of KL. The discrete uniform distribution with $N_\mathrm{e}+1$ possible values is $U(r_j) = (N_\mathrm{e}+1)^{-1}$ at each $r_j$. We then get the KL as
\begin{equation}
    \mathrm{KL}(U,p_r) 
    = \sum_{j=0}^{N_\mathrm{e}} U(r_j) \ln \left ( \frac{U(r_j)}{p_r(r_j)} \right ) 
    = -\frac{1}{N_\mathrm{e}+1} \sum_{j=0}^{N_\mathrm{e}} \ln \left ( \left (N_\mathrm{e}+1 \right ) p_r(r_j) \right ).
\end{equation}

\subsubsection{Fractions skill score} \label{sect:fss}

In precipitation forecasts, one often wants to predict whether the precipitation exceeds a certain threshold level $T$. Using the notation of the previous sections, we define the occurrence of such events as binary variables:
\begin{eqnarray}
S_i & = & H(y_i-T) \\
\hat S_{i,k} & = & H(\hat y_{i,k}-T)
\end{eqnarray}
where $S_i$ is the observed occurrence of the threshold-exceeding event at point $i$ and $\hat S_{i,k}$ is the occurrence in the forecast at $i$ in the ensemble member $k$.

The FSS \cite{Roberts2008FSS} is based on the notion that predicting the location of an event wrong by a short distance should be penalized less than mispredicting it by a long distance. Most scores such as root-mean-square error (RMSE), MAE or the critical success index (CSI; also known as the threat score or the intersection-over-union score) do not have this property; they penalize incorrect predictions equally regardless of whether or not there is a correct prediction nearby. 

In the calculation of FSS, one first defines the \emph{fraction} of events in a neighborhood $V$ of points as:
\begin{eqnarray}
M_V & = & \frac{1}{|V|} \sum_{i \in V} S_i \\
\hat M_V & = & \frac{1}{N_\mathrm{e}|V|} \sum_{i \in V} \sum_{k=1}^{N_\mathrm{e}} \hat S_{i,k}
\end{eqnarray}
where $|V|$ denotes the number of points in $V$. In the definition of $\hat M_V$, we use the generalization of \cite{Duc2013FSSEnsemble} to ensemble forecasts. To calculate FSS at a given spatial scale $n$, we define $W_{(n)}$ as the set of all square neighborhoods of $n \times n$ size. This follows common practice and simplifies calculation; alternatively one can use, for instance, circular neighborhoods. The fractional Brier score $\mathrm{FBS}_\mathrm{(n)}$ and the reference FBS (i.e. the FBS of a skilless forecast) $\mathrm{FBS}_\mathrm{(n),ref}$ for the scale $n$ are given by
\begin{eqnarray}
\mathrm{FBS}_\mathrm{(n)} &=& \frac{1}{\left | W_{(n)} \right |} \sum_{V \in W_{(n)}} \left ( \hat M_V - M_V \right )^2 \\
\mathrm{FBS}_\mathrm{(n),ref} &=& \frac{1}{\left | W_{(n)} \right |} \sum_{V \in W_{(n)}} \hat M_V^2 + M_V^2 .
\end{eqnarray}
Finally, the FSS for scale $n$ is
\begin{equation}
\mathrm{FSS}_\mathrm{(n)} = 1 - \frac{\mathrm{FBS}_\mathrm{(n)}}{\mathrm{FBS}_\mathrm{(n),ref}} .
\end{equation}
FSS is $1$ for an ideal forecast and $0$ for a skilless forecast.

\subsection{Benchmarks}

\subsubsection{Deep Generative Models of Radar}

DGMR \cite{Ravuri2021GenerativePrecipitation} represents the current state of the art in generative nowcasting. It is a GAN generator that was trained with a GAN hinge loss combined with a regularization loss that encourages the ensemble mean of the generated precipitation fields to match the true precipitation amount. The generator is built using convolutional gated recurrent unit (ConvGRU) layers organized in a U-Net-like structure, while the discriminator is split into separate spatial and temporal discriminators that both use convolutional layers. The GAN was trained with a dataset of radar-measured precipitation from the UK Met Office RadarNet4 network of C-band polarimetric radars.

The DGMR authors have made a saved model available, and we use it as our main point of comparison to LDCast. The inputs are compatible with our model, as the available model is trained for $256 \times 256$ pixel inputs at $1\ \mathrm{km}$ spatial and $5\ \mathrm{min}$ temporal resolution. DGMR produces an output up to $90\ \mathrm{min}$ to the future. Because of the spatiotemporal latent-space encoding our model must produce forecasts of a multiple of $4$ time steps ($20\ \mathrm{min}$), so we trained it to predict up to $100\ \mathrm{min}$ into the future and truncated the results at $90\ \mathrm{min}$ when computing scores that are compared directly to DGMR.

\subsubsection{PySTEPS}

PySTEPS \cite{Pulkkinen2019Pysteps} is a nowcasting library that implements the STEPS algorithm for stochastic ensemble nowcasting. We include PySTEPS in the comparisons presented in this paper as a state-of-the-art non-ML-based method. Extensive comparisons between PySTEPS and DGMR can also be found in \cite{Ravuri2021GenerativePrecipitation}. We used PySTEPS following the STEPS example on the PySTEPS website \footnote{\url{https://pysteps.readthedocs.io/en/stable/auto_examples/plot_steps_nowcast.html}}. 

We produced an output of zero rainfall for PySTEPS whenever the input was all zeros. We also found that occasional samples in our datasets caused the PySTEPS processing to fail. Examination of these cases showed that the problems occurred with very low rain rates, so we produced an output of all zero precipitation whenever this happened.

\section*{Data availability} \label{sect:dataavailability}

The pretrained models and the training and evaluation datasets can be found at \cite{Leinonen2023LDCastModelsData}.

\section*{Code availability} \label{sect:codeavailability}

The code for replicating the results can be found at \url{https://github.com/MeteoSwiss/ldcast}. 

The saved DGMR generator model can be found \url{https://github.com/deepmind/deepmind-research/tree/master/nowcasting}. The PySTEPS library website is \url{https://pysteps.github.io/}; PySTEPS can also be installed through many Python package managers.

\bibliographystyle{naturemag}  
\bibliography{journalabrv,ldmnowcast}  

\begin{thebibliography}{10}
\expandafter\ifx\csname url\endcsname\relax
  \def\url#1{\texttt{#1}}\fi
\expandafter\ifx\csname urlprefix\endcsname\relax\def\urlprefix{URL }\fi
\providecommand{\bibinfo}[2]{#2}
\providecommand{\eprint}[2][]{\url{#2}}

\bibitem{Surcel2015ScaleDependence}
\bibinfo{author}{Surcel, M.}, \bibinfo{author}{Zawadzki, I.} \&
  \bibinfo{author}{Yau, M.~K.}
\newblock \bibinfo{title}{A study on the scale dependence of the predictability
  of precipitation patterns}.
\newblock \emph{\bibinfo{journal}{J.\ Atmos.\ Sci.}}
  \textbf{\bibinfo{volume}{72}}, \bibinfo{pages}{216--235}
  (\bibinfo{year}{2015}).
\newblock \urlprefix\url{https://doi.org/10.1175/JAS-D-14-0071.1}.

\bibitem{Sun2014NWPPrecip}
\bibinfo{author}{Sun, J.} \emph{et~al.}
\newblock \bibinfo{title}{Use of {NWP} for nowcasting convective precipitation:
  Recent progress and challenges}.
\newblock \emph{\bibinfo{journal}{Bull.\ Amer.\ Meteor.\ Soc.}}
  \textbf{\bibinfo{volume}{95}}, \bibinfo{pages}{409--426}
  (\bibinfo{year}{2014}).

\bibitem{Bellon1978SHARP}
\bibinfo{author}{Bellon, A.} \& \bibinfo{author}{Austin, G.~L.}
\newblock \bibinfo{title}{The evaluation of two years of real-time operation of
  a short-term precipitation forecasting procedure {(SHARP)}}.
\newblock \emph{\bibinfo{journal}{J.\ Appl.\ Meteor.}}
  \textbf{\bibinfo{volume}{17}}, \bibinfo{pages}{1778--1787}
  (\bibinfo{year}{1978}).
\newblock \urlprefix\url{http://www.jstor.org/stable/26178613}.

\bibitem{Germann2002ScaleDependence}
\bibinfo{author}{Germann, U.} \& \bibinfo{author}{Zawadzki, I.}
\newblock \bibinfo{title}{Scale-dependence of the predictability of
  precipitation from continental radar images. {Part I}: Description of the
  methodology}.
\newblock \emph{\bibinfo{journal}{Mon.\ Wea.\ Rev.}}
  \textbf{\bibinfo{volume}{130}}, \bibinfo{pages}{2859--2873}
  (\bibinfo{year}{2002}).
\newblock
  \urlprefix\url{https://doi.org/10.1175/1520-0493(2002)130<2859:SDOTPO>2.0.CO;2}.

\bibitem{Bowler2006STEPS}
\bibinfo{author}{Bowler, N.~E.}, \bibinfo{author}{Pierce, C.~E.} \&
  \bibinfo{author}{Seed, A.~W.}
\newblock \bibinfo{title}{{STEPS}: A probabilistic precipitation forecasting
  scheme which merges an extrapolation nowcast with downscaled {NWP}}.
\newblock \emph{\bibinfo{journal}{Quart.\ J.\ Roy.\ Meteor.\ Soc.}}
  \textbf{\bibinfo{volume}{132}}, \bibinfo{pages}{2127--2155}
  (\bibinfo{year}{2006}).
\newblock \urlprefix\url{https://doi.org/10.1256/qj.04.100}.

\bibitem{Sideris2020NowPrecip}
\bibinfo{author}{Sideris, I.~V.}, \bibinfo{author}{Foresti, L.},
  \bibinfo{author}{Nerini, D.} \& \bibinfo{author}{Germann, U.}
\newblock \bibinfo{title}{{NowPrecip}: localized precipitation nowcasting in
  the complex terrain of {Switzerland}}.
\newblock \emph{\bibinfo{journal}{Quart.\ J.\ Roy.\ Meteor.\ Soc.}}
  \textbf{\bibinfo{volume}{146}}, \bibinfo{pages}{1768--1800}
  (\bibinfo{year}{2020}).
\newblock \urlprefix\url{https://doi.org/10.1002/qj.3766}.

\bibitem{Panziera2011NowcastingOrographic}
\bibinfo{author}{Panziera, L.}, \bibinfo{author}{Germann, U.},
  \bibinfo{author}{Gabella, M.} \& \bibinfo{author}{Mandapaka, P.~V.}
\newblock \bibinfo{title}{{NORA}–nowcasting of orographic rainfall by means
  of analogues}.
\newblock \emph{\bibinfo{journal}{Quart.\ J.\ Roy.\ Meteor.\ Soc.}}
  \textbf{\bibinfo{volume}{137}}, \bibinfo{pages}{2106--2123}
  (\bibinfo{year}{2011}).
\newblock \urlprefix\url{https://doi.org/10.1002/qj.878}.

\bibitem{Foresti2018AnalysisOrographic}
\bibinfo{author}{Foresti, L.}, \bibinfo{author}{Sideris, I.~V.},
  \bibinfo{author}{Panziera, L.}, \bibinfo{author}{Nerini, D.} \&
  \bibinfo{author}{Germann, U.}
\newblock \bibinfo{title}{A 10-year radar-based analysis of orographic
  precipitation growth and decay patterns over the {Swiss} {Alpine} region}.
\newblock \emph{\bibinfo{journal}{Quart.\ J.\ Roy.\ Meteor.\ Soc.}}
  \textbf{\bibinfo{volume}{144}}, \bibinfo{pages}{2277--2301}
  (\bibinfo{year}{2018}).
\newblock \urlprefix\url{https://doi.org/10.1002/qj.3364}.

\bibitem{Seed2013STEPS}
\bibinfo{author}{Seed, A.~W.}, \bibinfo{author}{Pierce, C.~E.} \&
  \bibinfo{author}{Norman, K.}
\newblock \bibinfo{title}{Formulation and evaluation of a scale
  decomposition-based stochastic precipitation nowcast scheme}.
\newblock \emph{\bibinfo{journal}{Water Resour. Res.}}
  \textbf{\bibinfo{volume}{49}}, \bibinfo{pages}{6624--6641}
  (\bibinfo{year}{2013}).
\newblock \urlprefix\url{https://doi.org/10.1002/wrcr.20536}.

\bibitem{Pulkkinen2019Pysteps}
\bibinfo{author}{Pulkkinen, S.} \emph{et~al.}
\newblock \bibinfo{title}{Pysteps: an open-source {Python} library for
  probabilistic precipitation nowcasting (v1.0)}.
\newblock \emph{\bibinfo{journal}{Geosci. Model Dev.}}
  \textbf{\bibinfo{volume}{12}}, \bibinfo{pages}{4185--4219}
  (\bibinfo{year}{2019}).
\newblock \urlprefix\url{https://doi.org/10.5194/gmd-12-4185-2019/}.

\bibitem{Shi2015DeepNowcast}
\bibinfo{author}{Shi, X.} \emph{et~al.}
\newblock \bibinfo{title}{Convolutional {LSTM} network: A machine learning
  approach for precipitation nowcasting}.
\newblock In \bibinfo{editor}{Cortes, C.}, \bibinfo{editor}{Lawrence, N.},
  \bibinfo{editor}{Lee, D.}, \bibinfo{editor}{Sugiyama, M.} \&
  \bibinfo{editor}{Garnett, R.} (eds.) \emph{\bibinfo{booktitle}{Advances in
  Neural Information Processing Systems}}, vol.~\bibinfo{volume}{28}
  (\bibinfo{publisher}{Curran Associates, Inc.}, \bibinfo{year}{2015}).
\newblock
  \urlprefix\url{https://proceedings.neurips.cc/paper/2015/file/07563a3fe3bbe7e3ba84431ad9d055af-Paper.pdf}.

\bibitem{Agrawal2019Nowcast}
\bibinfo{author}{Agrawal, S.} \emph{et~al.}
\newblock \bibinfo{title}{Machine learning for precipitation nowcasting from
  radar images} (\bibinfo{year}{2019}).
\newblock \urlprefix\url{https://arxiv.org/abs/1912.12132}.

\bibitem{Ayzel2020RainNet}
\bibinfo{author}{Ayzel, G.}, \bibinfo{author}{Scheffer, T.} \&
  \bibinfo{author}{Heistermann, M.}
\newblock \bibinfo{title}{{RainNet} v1.0: a~convolutional neural network for
  radar-based precipitation nowcasting}.
\newblock \emph{\bibinfo{journal}{Geosci. Model Dev.}}
  \textbf{\bibinfo{volume}{13}}, \bibinfo{pages}{2631--2644}
  (\bibinfo{year}{2020}).
\newblock \urlprefix\url{https://doi.org/10.5194/gmd-13-2631-2020/}.

\bibitem{Franch2020Nowcast}
\bibinfo{author}{Franch, G.} \emph{et~al.}
\newblock \bibinfo{title}{Precipitation nowcasting with orographic enhanced
  stacked generalization: Improving deep learning predictions on extreme
  events}.
\newblock \emph{\bibinfo{journal}{Atmosphere}} \textbf{\bibinfo{volume}{11}}
  (\bibinfo{year}{2020}).
\newblock \urlprefix\url{https://doi.org/10.3390/atmos11030267}.

\bibitem{Goodfellow2020GAN}
\bibinfo{author}{Goodfellow, I.} \emph{et~al.}
\newblock \bibinfo{title}{Generative adversarial networks}.
\newblock \emph{\bibinfo{journal}{Commun. ACM}} \textbf{\bibinfo{volume}{63}},
  \bibinfo{pages}{139--144} (\bibinfo{year}{2020}).
\newblock \urlprefix\url{https://doi.org/10.1145/3422622}.

\bibitem{Leinonen2020Downscaling}
\bibinfo{author}{Leinonen, J.}, \bibinfo{author}{Nerini, D.} \&
  \bibinfo{author}{Berne, A.}
\newblock \bibinfo{title}{Stochastic super-resolution for downscaling
  time-evolving atmospheric fields with a generative adversarial network}.
\newblock \emph{\bibinfo{journal}{{IEEE} Trans. Geosci. Remote Sens.}}
  \textbf{\bibinfo{volume}{59}}, \bibinfo{pages}{7211--7223}
  (\bibinfo{year}{2021}).
\newblock \urlprefix\url{https://doi.org/10.1109/TGRS.2020.3032790}.

\bibitem{Price2022Resolution}
\bibinfo{author}{Price, I.} \& \bibinfo{author}{Rasp, S.}
\newblock \bibinfo{title}{Increasing the accuracy and resolution of
  precipitation forecasts using deep generative models}.
\newblock In \bibinfo{editor}{Camps-Valls, G.}, \bibinfo{editor}{Ruiz, F.
  J.~R.} \& \bibinfo{editor}{Valera, I.} (eds.)
  \emph{\bibinfo{booktitle}{Proceedings of The 25th International Conference on
  Artificial Intelligence and Statistics}}, vol. \bibinfo{volume}{151} of
  \emph{\bibinfo{series}{Proceedings of Machine Learning Research}},
  \bibinfo{pages}{10555--10571} (\bibinfo{publisher}{PMLR},
  \bibinfo{year}{2022}).
\newblock \urlprefix\url{https://proceedings.mlr.press/v151/price22a.html}.

\bibitem{Harris2022Downscaling}
\bibinfo{author}{Harris, L.}, \bibinfo{author}{McRae, A. T.~T.},
  \bibinfo{author}{Chantry, M.}, \bibinfo{author}{Dueben, P.~D.} \&
  \bibinfo{author}{Palmer, T.~N.}
\newblock \bibinfo{title}{A generative deep learning approach to stochastic
  downscaling of precipitation forecasts}.
\newblock \emph{\bibinfo{journal}{J. Adv. Model. Earth Sys.}}
  \textbf{\bibinfo{volume}{14}}, \bibinfo{pages}{e2022MS003120}
  (\bibinfo{year}{2022}).
\newblock \urlprefix\url{https://doi.org/10.1029/2022MS003120}.

\bibitem{Hayatbini2019PERSIANNGAN}
\bibinfo{author}{Hayatbini, N.} \emph{et~al.}
\newblock \bibinfo{title}{Conditional generative adversarial networks {(cGANs)}
  for near real-time precipitation estimation from multispectral {GOES-16}
  satellite imageries --- {PERSIANN-cGAN}}.
\newblock \emph{\bibinfo{journal}{Remote Sens.}} \textbf{\bibinfo{volume}{11}}
  (\bibinfo{year}{2019}).
\newblock \urlprefix\url{https://doi.org/10.3390/rs11192193}.

\bibitem{Wang2021PrecipGAN}
\bibinfo{author}{Wang, C.}, \bibinfo{author}{Tang, G.} \&
  \bibinfo{author}{Gentine, P.}
\newblock \bibinfo{title}{{PrecipGAN}: Merging microwave and infrared data for
  satellite precipitation estimation using generative adversarial network}.
\newblock \emph{\bibinfo{journal}{Geophys.\ Res.\ Lett.}}
  \textbf{\bibinfo{volume}{48}}, \bibinfo{pages}{e2020GL092032}
  (\bibinfo{year}{2021}).
\newblock \urlprefix\url{https://doi.org/10.1029/2020GL092032}.

\bibitem{Scher2021Disaggregation}
\bibinfo{author}{Scher, S.} \& \bibinfo{author}{Pe{\ss}enteiner, S.}
\newblock \bibinfo{title}{Technical note: Temporal disaggregation of spatial
  rainfall fields with generative adversarial networks}.
\newblock \emph{\bibinfo{journal}{Hydrol. Earth Syst. Sci.}}
  \textbf{\bibinfo{volume}{25}}, \bibinfo{pages}{3207--3225}
  (\bibinfo{year}{2021}).
\newblock \urlprefix\url{https://doi.org/10.5194/hess-25-3207-2021}.

\bibitem{Ravuri2021GenerativePrecipitation}
\bibinfo{author}{Ravuri, S.} \emph{et~al.}
\newblock \bibinfo{title}{Skilful precipitation nowcasting using deep
  generative models of radar}.
\newblock \emph{\bibinfo{journal}{Nature}} \textbf{\bibinfo{volume}{597}},
  \bibinfo{pages}{672--677} (\bibinfo{year}{2021}).
\newblock \urlprefix\url{https://doi.org/10.1038/s41586-021-03854-z/}.

\bibitem{Mescheder2018GANTraining}
\bibinfo{author}{Mescheder, L.}, \bibinfo{author}{Geiger, A.} \&
  \bibinfo{author}{Nowozin, S.}
\newblock \bibinfo{title}{Which training methods for {GAN}s do actually
  converge?}
\newblock In \bibinfo{editor}{Dy, J.} \& \bibinfo{editor}{Krause, A.} (eds.)
  \emph{\bibinfo{booktitle}{Proceedings of the 35th International Conference on
  Machine Learning}}, vol.~\bibinfo{volume}{80} of
  \emph{\bibinfo{series}{Proceedings of Machine Learning Research}},
  \bibinfo{pages}{3481--3490} (\bibinfo{publisher}{PMLR},
  \bibinfo{year}{2018}).
\newblock \urlprefix\url{https://proceedings.mlr.press/v80/mescheder18a.html}.

\bibitem{Bau2019ModeCollapse}
\bibinfo{author}{Bau, D.} \emph{et~al.}
\newblock \bibinfo{title}{Seeing what a {GAN} cannot generate}.
\newblock In \emph{\bibinfo{booktitle}{Proceedings of the IEEE/CVF
  International Conference on Computer Vision (ICCV)}} (\bibinfo{year}{2019}).
\newblock \urlprefix\url{https://doi.org/10.1109/ICCV.2019.00460}.

\bibitem{Song2019GenerativeGradients}
\bibinfo{author}{Song, Y.} \& \bibinfo{author}{Ermon, S.}
\newblock \bibinfo{title}{Generative modeling by estimating gradients of the
  data distribution}.
\newblock In \bibinfo{editor}{Wallach, H.} \emph{et~al.} (eds.)
  \emph{\bibinfo{booktitle}{Advances in Neural Information Processing
  Systems}}, vol.~\bibinfo{volume}{32} (\bibinfo{publisher}{Curran Associates,
  Inc.}, \bibinfo{year}{2019}).
\newblock
  \urlprefix\url{https://proceedings.neurips.cc/paper/2019/file/3001ef257407d5a371a96dcd947c7d93-Paper.pdf}.

\bibitem{Song2020ImprovedScoreBased}
\bibinfo{author}{Song, Y.} \& \bibinfo{author}{Ermon, S.}
\newblock \bibinfo{title}{Improved techniques for training score-based
  generative models}.
\newblock In \bibinfo{editor}{Larochelle, H.}, \bibinfo{editor}{Ranzato, M.},
  \bibinfo{editor}{Hadsell, R.}, \bibinfo{editor}{Balcan, M.} \&
  \bibinfo{editor}{Lin, H.} (eds.) \emph{\bibinfo{booktitle}{Advances in Neural
  Information Processing Systems}}, vol.~\bibinfo{volume}{33},
  \bibinfo{pages}{12438--12448} (\bibinfo{publisher}{Curran Associates, Inc.},
  \bibinfo{year}{2020}).
\newblock
  \urlprefix\url{https://proceedings.neurips.cc/paper/2020/file/92c3b916311a5517d9290576e3ea37ad-Paper.pdf}.

\bibitem{Dhariwal2021DiffusionGAN}
\bibinfo{author}{Dhariwal, P.} \& \bibinfo{author}{Nichol, A.}
\newblock \bibinfo{title}{Diffusion models beat {GANs} on image synthesis}.
\newblock In \bibinfo{editor}{Ranzato, M.}, \bibinfo{editor}{Beygelzimer, A.},
  \bibinfo{editor}{Dauphin, Y.}, \bibinfo{editor}{Liang, P.} \&
  \bibinfo{editor}{Vaughan, J.~W.} (eds.) \emph{\bibinfo{booktitle}{Advances in
  Neural Information Processing Systems}}, vol.~\bibinfo{volume}{34},
  \bibinfo{pages}{8780--8794} (\bibinfo{publisher}{Curran Associates, Inc.},
  \bibinfo{year}{2021}).
\newblock
  \urlprefix\url{https://proceedings.neurips.cc/paper/2021/file/49ad23d1ec9fa4bd8d77d02681df5cfa-Paper.pdf}.

\bibitem{Saharia2022Palette}
\bibinfo{author}{Saharia, C.} \emph{et~al.}
\newblock \bibinfo{title}{Palette: Image-to-image diffusion models}.
\newblock In \emph{\bibinfo{booktitle}{ACM SIGGRAPH 2022 Conference
  Proceedings}}, SIGGRAPH '22 (\bibinfo{publisher}{Association for Computing
  Machinery}, \bibinfo{address}{New York, NY, USA}, \bibinfo{year}{2022}).
\newblock \urlprefix\url{https://doi.org/10.1145/3528233.3530757}.

\bibitem{Ramesh2022DALLE2}
\bibinfo{author}{Ramesh, A.}, \bibinfo{author}{Dhariwal, P.},
  \bibinfo{author}{Nichol, A.}, \bibinfo{author}{Chu, C.} \&
  \bibinfo{author}{Chen, M.}
\newblock \bibinfo{title}{Hierarchical text-conditional image generation with
  {CLIP} latents} (\bibinfo{year}{2022}).
\newblock \urlprefix\url{https://arxiv.org/abs/2204.06125}.
\newblock \eprint{2204.06125}.

\bibitem{Saharia2022Imagen}
\bibinfo{author}{Saharia, C.} \emph{et~al.}
\newblock \bibinfo{title}{Photorealistic text-to-image diffusion models with
  deep language understanding}.
\newblock In \bibinfo{editor}{Oh, A.~H.}, \bibinfo{editor}{Agarwal, A.},
  \bibinfo{editor}{Belgrave, D.} \& \bibinfo{editor}{Cho, K.} (eds.)
  \emph{\bibinfo{booktitle}{Advances in Neural Information Processing Systems}}
  (\bibinfo{year}{2022}).
\newblock \urlprefix\url{https://openreview.net/forum?id=08Yk-n5l2Al}.

\bibitem{Li2022SRDiff}
\bibinfo{author}{Li, H.} \emph{et~al.}
\newblock \bibinfo{title}{{SRDiff}: Single image super-resolution with
  diffusion probabilistic models}.
\newblock \emph{\bibinfo{journal}{Neurocomputing}}
  \textbf{\bibinfo{volume}{479}}, \bibinfo{pages}{47--59}
  (\bibinfo{year}{2022}).
\newblock \urlprefix\url{https://doi.org/10.1016/j.neucom.2022.01.029}.

\bibitem{Ho2020DDPM}
\bibinfo{author}{Ho, J.}, \bibinfo{author}{Jain, A.} \&
  \bibinfo{author}{Abbeel, P.}
\newblock \bibinfo{title}{Denoising diffusion probabilistic models}.
\newblock In \bibinfo{editor}{Larochelle, H.}, \bibinfo{editor}{Ranzato, M.},
  \bibinfo{editor}{Hadsell, R.}, \bibinfo{editor}{Balcan, M.} \&
  \bibinfo{editor}{Lin, H.} (eds.) \emph{\bibinfo{booktitle}{Advances in Neural
  Information Processing Systems}}, vol.~\bibinfo{volume}{33},
  \bibinfo{pages}{6840--6851} (\bibinfo{publisher}{Curran Associates, Inc.},
  \bibinfo{year}{2020}).
\newblock
  \urlprefix\url{https://proceedings.neurips.cc/paper/2020/file/4c5bcfec8584af0d967f1ab10179ca4b-Paper.pdf}.

\bibitem{Song2021DDIM}
\bibinfo{author}{Song, J.}, \bibinfo{author}{Meng, C.} \&
  \bibinfo{author}{Ermon, S.}
\newblock \bibinfo{title}{Denoising diffusion implicit models}.
\newblock In \emph{\bibinfo{booktitle}{International Conference on Learning
  Representations}} (\bibinfo{year}{2021}).
\newblock \urlprefix\url{https://openreview.net/forum?id=St1giarCHLP}.

\bibitem{Liu2022PLMS}
\bibinfo{author}{Liu, L.}, \bibinfo{author}{Ren, Y.}, \bibinfo{author}{Lin, Z.}
  \& \bibinfo{author}{Zhao, Z.}
\newblock \bibinfo{title}{Pseudo numerical methods for diffusion models on
  manifolds}.
\newblock In \emph{\bibinfo{booktitle}{International Conference on Learning
  Representations}} (\bibinfo{year}{2022}).
\newblock \urlprefix\url{https://openreview.net/forum?id=PlKWVd2yBkY}.

\bibitem{Addison2022Emulation}
\bibinfo{author}{Addison, H.}, \bibinfo{author}{Kendon, E.},
  \bibinfo{author}{Ravuri, S.}, \bibinfo{author}{Aitchison, L.} \&
  \bibinfo{author}{Watson, P.~A.}
\newblock \bibinfo{title}{Machine learning emulation of a local-scale {UK}
  climate model} (\bibinfo{year}{2022}).
\newblock \urlprefix\url{https://arxiv.org/abs/2211.16116}.

\bibitem{Rombach2022LatentDiffusion}
\bibinfo{author}{Rombach, R.}, \bibinfo{author}{Blattmann, A.},
  \bibinfo{author}{Lorenz, D.}, \bibinfo{author}{Esser, P.} \&
  \bibinfo{author}{Ommer, B.}
\newblock \bibinfo{title}{High-resolution image synthesis with latent diffusion
  models}.
\newblock In \emph{\bibinfo{booktitle}{Proceedings of the IEEE/CVF Conference
  on Computer Vision and Pattern Recognition (CVPR)}},
  \bibinfo{pages}{10684--10695} (\bibinfo{year}{2022}).
\newblock \urlprefix\url{https://arxiv.org/abs/2112.10752}.

\bibitem{Guibas2022AFNO}
\bibinfo{author}{Guibas, J.} \emph{et~al.}
\newblock \bibinfo{title}{Efficient token mixing for transformers via {Adaptive
  Fourier Neural Operators}}.
\newblock In \emph{\bibinfo{booktitle}{International Conference on Learning
  Representations}} (\bibinfo{year}{2022}).
\newblock \urlprefix\url{https://arxiv.org/abs/2111.13587}.

\bibitem{Pathak2022FourCastNet}
\bibinfo{author}{Pathak, J.} \emph{et~al.}
\newblock \bibinfo{title}{{FourCastNet}: A global data-driven high-resolution
  weather model using {Adaptive Fourier Neural Operators}}
  (\bibinfo{year}{2022}).
\newblock \urlprefix\url{https://arxiv.org/abs/2202.11214}.
\newblock \eprint{2202.11214}.

\bibitem{Esser2021Taming}
\bibinfo{author}{Esser, P.}, \bibinfo{author}{Rombach, R.} \&
  \bibinfo{author}{Ommer, B.}
\newblock \bibinfo{title}{Taming transformers for high-resolution image
  synthesis}.
\newblock In \emph{\bibinfo{booktitle}{Proceedings of the IEEE/CVF Conference
  on Computer Vision and Pattern Recognition (CVPR)}},
  \bibinfo{pages}{12873--12883} (\bibinfo{year}{2021}).
\newblock \urlprefix\url{https://doi.org/10.1109/CVPR46437.2021.01268}.

\bibitem{Leinonen2022Lightning}
\bibinfo{author}{Leinonen, J.}, \bibinfo{author}{Hamann, U.} \&
  \bibinfo{author}{Germann, U.}
\newblock \bibinfo{title}{Seamless lightning nowcasting with
  recurrent-convolutional deep learning}.
\newblock \emph{\bibinfo{journal}{Artif. Intell. Earth Syst.}}
  \textbf{\bibinfo{volume}{1}}, \bibinfo{pages}{e220043}
  (\bibinfo{year}{2022}).
\newblock \urlprefix\url{https://doi.org/10.1175/AIES-D-22-0043.1}.

\bibitem{Germann2006RadarMountainous}
\bibinfo{author}{Germann, U.}, \bibinfo{author}{Galli, G.},
  \bibinfo{author}{Boscacci, M.} \& \bibinfo{author}{Bolliger, M.}
\newblock \bibinfo{title}{Radar precipitation measurement in a mountainous
  region}.
\newblock \emph{\bibinfo{journal}{Quart.\ J.\ Roy.\ Meteor.\ Soc.}}
  \textbf{\bibinfo{volume}{132}}, \bibinfo{pages}{1669--1692}
  (\bibinfo{year}{2006}).
\newblock \urlprefix\url{https://doi.org/10.1256/qj.05.190}.

\bibitem{Germann2016SwissRadar}
\bibinfo{author}{Germann, U.}, \bibinfo{author}{Boscacci, M.},
  \bibinfo{author}{Gabella, M.} \& \bibinfo{author}{Schneebeli, M.}
\newblock \bibinfo{title}{Weather radar in {Switzerland}}.
\newblock In \bibinfo{editor}{Willemse, S.} \& \bibinfo{editor}{Furger, M.}
  (eds.) \emph{\bibinfo{booktitle}{From weather observations to atmospheric and
  climate science in {Switzerland}: {Celebrating} 100 years of the {Swiss
  Society for Meteorology}}}, chap.~\bibinfo{chapter}{9}
  (\bibinfo{publisher}{Vdf Hochschulverlag AG an der ETH Zürich},
  \bibinfo{address}{Zürich, Switzerland}, \bibinfo{year}{2016}).

\bibitem{Stephan2008DWDRadar}
\bibinfo{author}{Stephan, K.}, \bibinfo{author}{Klink, S.} \&
  \bibinfo{author}{Schraff, C.}
\newblock \bibinfo{title}{Assimilation of radar-derived rain rates into the
  convective-scale model {COSMO-DE} at {DWD}}.
\newblock \emph{\bibinfo{journal}{Quart.\ J.\ Roy.\ Meteor.\ Soc.}}
  \textbf{\bibinfo{volume}{134}}, \bibinfo{pages}{1315--1326}
  (\bibinfo{year}{2008}).
\newblock \urlprefix\url{https://doi.org/10.1002/qj.269}.

\bibitem{He2016ResNet}
\bibinfo{author}{He, K.}, \bibinfo{author}{Zhang, X.}, \bibinfo{author}{Ren,
  S.} \& \bibinfo{author}{Sun, J.}
\newblock \bibinfo{title}{Deep residual learning for image recognition}.
\newblock In \emph{\bibinfo{booktitle}{The IEEE Conference on Computer Vision
  and Pattern Recognition (CVPR)}} (\bibinfo{year}{2016}).
\newblock \urlprefix\url{https://doi.org/10.1109/CVPR.2016.90}.

\bibitem{Vaswani2017Attention}
\bibinfo{author}{Vaswani, A.} \emph{et~al.}
\newblock \bibinfo{title}{Attention is all you need}.
\newblock In \bibinfo{editor}{Guyon, I.} \emph{et~al.} (eds.)
  \emph{\bibinfo{booktitle}{Advances in Neural Information Processing
  Systems}}, vol.~\bibinfo{volume}{30} (\bibinfo{publisher}{Curran Associates,
  Inc.}, \bibinfo{year}{2017}).
\newblock
  \urlprefix\url{https://proceedings.neurips.cc/paper/2017/file/3f5ee243547dee91fbd053c1c4a845aa-Paper.pdf}.

\bibitem{Leinonen2023ThunderstormHazard}
\bibinfo{author}{Leinonen, J.}, \bibinfo{author}{Hamann, U.},
  \bibinfo{author}{Sideris, I.~V.} \& \bibinfo{author}{Germann, U.}
\newblock \bibinfo{title}{Thunderstorm nowcasting with deep learning: a
  multi-hazard data fusion model}.
\newblock \emph{\bibinfo{journal}{Geophys.\ Res.\ Lett.}}
  (\bibinfo{year}{2023}).
\newblock \urlprefix\url{https://doi.org/10.1029/2022GL101626}.
\newblock \bibinfo{note}{Accepted for publication}.

\bibitem{Loshchilov2018AdamW}
\bibinfo{author}{Loshchilov, I.} \& \bibinfo{author}{Hutter, F.}
\newblock \bibinfo{title}{Decoupled weight decay regularization}.
\newblock In \emph{\bibinfo{booktitle}{International Conference on Learning
  Representations}} (\bibinfo{year}{2019}).
\newblock \urlprefix\url{https://openreview.net/forum?id=Bkg6RiCqY7}.

\bibitem{Gneiting2007ProperScoring}
\bibinfo{author}{Gneiting, T.} \& \bibinfo{author}{Raftery, A.~E.}
\newblock \bibinfo{title}{Strictly proper scoring rules, prediction, and
  estimation}.
\newblock \emph{\bibinfo{journal}{J. Am. Stat. Assoc.}}
  \textbf{\bibinfo{volume}{102}}, \bibinfo{pages}{359--378}
  (\bibinfo{year}{2007}).
\newblock \urlprefix\url{https://doi.org/10.1198/016214506000001437}.

\bibitem{Candille2006PredictionEvaluation}
\bibinfo{author}{Candille, G.} \& \bibinfo{author}{Talagrand, O.}
\newblock \bibinfo{title}{Evaluation of probabilistic prediction systems for a
  scalar variable}.
\newblock \emph{\bibinfo{journal}{Quart.\ J.\ Roy.\ Meteor.\ Soc.}}
  \textbf{\bibinfo{volume}{131}}, \bibinfo{pages}{2131--2150}
  (\bibinfo{year}{2005}).
\newblock \urlprefix\url{https://doi.org/10.1256/qj.04.71}.

\bibitem{Roberts2008FSS}
\bibinfo{author}{Roberts, N.~M.} \& \bibinfo{author}{Lean, H.~W.}
\newblock \bibinfo{title}{Scale-selective verification of rainfall
  accumulations from high-resolution forecasts of convective events}.
\newblock \emph{\bibinfo{journal}{Mon.\ Wea.\ Rev.}}
  \textbf{\bibinfo{volume}{136}}, \bibinfo{pages}{78--97}
  (\bibinfo{year}{2008}).
\newblock \urlprefix\url{https://doi.org/10.1175/2007MWR2123.1}.

\bibitem{Duc2013FSSEnsemble}
\bibinfo{author}{Duc, L.}, \bibinfo{author}{Saito, K.} \&
  \bibinfo{author}{Seko, H.}
\newblock \bibinfo{title}{Spatial-temporal fractions verification for
  high-resolution ensemble forecasts}.
\newblock \emph{\bibinfo{journal}{Tellus A: Dyn. Meteorol. Oceanogr.}}
  \textbf{\bibinfo{volume}{65}}, \bibinfo{pages}{18171} (\bibinfo{year}{2013}).
\newblock \urlprefix\url{https://doi.org/10.3402/tellusa.v65i0.18171}.

\bibitem{Leinonen2023LDCastModelsData}
\bibinfo{author}{Leinonen, J.}, \bibinfo{author}{Hamann, U.},
  \bibinfo{author}{Nerini, D.}, \bibinfo{author}{Germann, U.} \&
  \bibinfo{author}{Franch, G.}
\newblock \bibinfo{title}{{Pretrained models and results for ``Latent diffusion
  models for generative precipitation nowcasting with accurate uncertainty
  quantification''}} (\bibinfo{year}{2023}).
\newblock \urlprefix\url{https://doi.org/10.5281/zenodo.7780914}.

\end{thebibliography}

\section*{Acknowledgments}

We thank Markus Schultze and Ulrich Blahak of DWD for providing the German radar dataset, and Nathalie Rombeek for feedback on the article.

\section*{Funding}

JL was supported by the fellowship ``Seamless Artificially Intelligent Thunderstorm Nowcasts'' from the European Organisation for the Exploitation of Meteorological Satellites (EUMETSAT). The hosting institution of this fellowship is MeteoSwiss in Switzerland.





\end{document}


\maketitle
\clearpage

\begin{table}
 \caption{The key architecture, training and evaluation hyperparameters used in the implementation of LDCast in this article.}
  \centering
  \begin{tabular}{ll}
    \toprule
    \multicolumn{2}{l}{\textbf{Autoencoder}} \\
    Up/downsampling channels & $64$ \\
    $2 \times 2 \times 2$ up/downsampling operations & $2$ \\
    Latent channels & $32$ \\
    \midrule
    \multicolumn{2}{l}{\textbf{Autoencoder training}} \\
    Loss & $L^1$ \\
    Initial learning rate (LR) & $10^{-3}$ \\
    KL regularization weight & $10^{-2}$ \\
    Epoch size (samples) & $1000$ \\
    Epochs w/o improvement before LR reduction & $3$ \\
    Epochs w/o improvement before early stopping & $6$ \\
    Batch size & $64$ \\
    \midrule
    \multicolumn{2}{l}{\textbf{Forecaster stack}} \\
    Channels & $128$ \\
    AFNO blocks (analysis) & $4$ \\
    AFNO blocks (forecast) & $4$ \\
    \midrule
    \multicolumn{2}{l}{\textbf{Denoiser stack}} \\
    Initial channels & $256$ \\
    U-Net Depth & $3$ \\
    U-Net channel multipliers & $1$, $2$, $4$ \\
    Number of ResNet blocks per level & $2$ \\
    \midrule
    \multicolumn{2}{l}{\textbf{LDM training}} \\
    Initial learning rate (LR) & $10^{-4}$ \\
    Epoch size (samples) & $1000$ \\
    Epochs w/o improvement before LR reduction & $3$ \\
    Epochs w/o improvement before early stopping & $6$ \\
    Batch size (initial) & $64$ \\
    Batch size (fine tuning, $256 \times 256$ samples) & $48$ \\
    \midrule
    \multicolumn{2}{l}{\textbf{Sample generation}} \\
    PLMS iterations & $50$ \\
    \bottomrule
  \end{tabular}
  \label{tab:table}
\end{table}